\documentclass[reprint,amsmath,amssymb,aps,floatfix,groupedaddress,superscriptaddress,nofootinbib,preprintnumbers]{revtex4-1}
\usepackage{float}
\usepackage[pdftex]{hyperref}
\usepackage{graphicx,color}
\usepackage{dcolumn}
\usepackage{bm}
\usepackage{here}
\usepackage{comment}
\usepackage[paperwidth=210mm,paperheight=297mm,centering,hmargin=1.5cm,vmargin=2.0cm]{geometry}
\usepackage{tabularx}
\newcolumntype{Y}{&gt;{\centering\arraybackslash}X} 
\hypersetup{
 setpagesize=false,
 bookmarksnumbered=true,%
 bookmarksopen=true,%
 colorlinks=true,%
 linkcolor=blue,
 citecolor=blue,
 urlcolor=blue,
}
\usepackage[left]{lineno}

\allowdisplaybreaks[1]

\begin{document}

\preprint{J-PARC-TH-0084}

\title{\bf More efficient formulas for efficiency correction of cumulants and\\ effect of using averaged efficiency}

\author{Toshihiro Nonaka}
\affiliation{Center\,for\,Integrated\,Research\,in\,Fundamental\,Science\,and\,Engineering,\,University\,of\,Tsukuba,\,Tsukuba,\,Ibaraki\,305,\,Japan}
\author{Masakiyo Kitazawa}
\affiliation{Department\,of\,Physics,\,Osaka\,University,\,Toyonaka,\,Osaka 560-0043,\,Japan}
\affiliation{J-PARC Branch,\,KEK Theory Center,\,Institute\,of\,Particle\,and\,Nuclear\,Studies,\,KEK,\,203-1,\,Shirakata,\,Tokai,\,Ibaraki,\,319-1106,\,Japan}
\author{ShinIchi Esumi}
\affiliation{Center\,for\,Integrated\,Research\,in\,Fundamental\,Science\,and\,Engineering,\,University\,of\,Tsukuba,\,Tsukuba,\,Ibaraki\,305,\,Japan}

\begin{abstract}
We derive formulas for the efficiency correction of cumulants
with many efficiency bins. The derivation of the formulas is simpler 
than the previously suggested method, but the numerical 
cost is drastically reduced from the na\"ive method.
From analytical and numerical analyses in simple toy models, 
We show that  use of the averaged efficiency in the efficiency correction 
might cause large deviations in some cases and should not be
used especially for high order cumulants.
These analyses show the importance of carrying out the efficiency 
correction without taking the average.

\end{abstract}

\pacs{Valid PACS appear here}
\maketitle
\newcommand{\ave}[1]{\ensuremath{\langle#1\rangle} }

\section{Introduction}
One of the major goals of heavy ion colliding experiments is to reveal the QCD phase structure. 
Event-by-event fluctuations of conserved quantities, such as net-baryon and net-charge distributions, 
have been proposed as experimental probes of the signal from the QCD critical point and phase transitions~\cite{fluctuation,Jeon:2000wg,susceptibility,correlation,Asakawa:2009aj,Friman}; 
see recent reviews Refs.~\cite{Asakawa:2015ybt,Luo:2017faz}.
The STAR experiment has measured the beam energy dependencies of the third and fourth order cumulants of net-proton and net-charge multiplicity distributions~\cite{net_proton,net_charge}.
In these studies enhancement and suppression of the cumulants at the low energy region are observed, which might be the signal of the critical point. 
However, there are still large statistical and systematic errors especially at low energy region.
The Beam Energy Scan Phase II program is planned at RHIC to accumulate much statistics.
In addition, the experimental group is also trying to measure the sixth order cumulant to find the signal of the phase transition~\cite{Lizhu}.

One of the experimental difficulties in the analyses of higher order cumulants
is concerned with finite detector efficiencies. We miss particles with some probability 
called efficiency, and the imperfect efficiency affects the shape of the event-by-event distributions and their cumulants~\cite{eff_kitazawa,eff_koch}.
The correction of this effect has been discussed in the literature~\cite{eff_kitazawa,eff_koch,eff_psd_volker,eff_xiaofeng,psd_Kitazawa,binomial_breaking,tsukuba_eff_separate}.
Moreover, in real detectors the efficiency often becomes non-uniform for detector acceptance due to many reasons, e.g., detector structures, or detector conditions for certain regions.
In this case, the non-uniformity of the efficiency should be taken into account in the correction.
Although the efficiency correction with non-uniform efficiencies are proposed in Refs.~\cite{eff_psd_volker,eff_xiaofeng,psd_Kitazawa}, 
these methods are difficult to apply to higher order cumulants. 
In the method proposed in Refs.~\cite{eff_psd_volker,eff_xiaofeng}, 
the numerical cost grows proportional to $M^m$, where $M$ is the number of efficiency bins and $m$ is the order of the cumulant.
Therefore, we cannot increase the numbers of $M$ and $m$ 
within a realistic CPU time.
In Ref.~\cite{psd_Kitazawa}, other method which drastically reduces 
the numerical cost has been proposed on the basis of cumulant expansion.
In this method, however, the derivation of the analytic formulas becomes  
complicated for higher order cumulants and it is difficult to apply 
the method to sixth order.
Therefore, an alternative efficient method for this problem is called for.

In the present paper, we propose a new method for the efficiency correction.
In this method, the analytic procedure is substantially simplified compared
to Ref.~\cite{psd_Kitazawa}. The numerical cost, however, is almost the same as 
that in Ref.~\cite{psd_Kitazawa}, and drastically smaller than those in
Refs.~\cite{eff_psd_volker,eff_xiaofeng}.

We also apply the formulas to practical problems in experiments.
Generally, acceptance uniformity of detectors can be violated due to various practical problems.
In this case, it is desirable to divide the efficiency bin into different acceptance regions
and apply the efficiency correction with the increased bins.
However, it is practically difficult to implement those corrections because of large numerical cost
especially for higher order cumulants. Then, one has to use a single averaged efficiency for these
acceptance regions.
In this paper, we study the effect of using the averaged efficiency in simple toy models. 
We show that  use of the averaged efficiency in the efficiency correction 
might cause large deviations in some cases and should not be
used especially for high order cumulants.

This paper is organized as follows. 
In Secs.~\ref{sec:single} and \ref{sec:multi}, we derive formulas for the efficiency correction with many efficiency bins. 
We first derive the result in a simple case with a single efficiency in Sec.~\ref{sec:single}, and then extend it to the multivariate case in Sec.~\ref{sec:multi}. 
In Sec.~\ref{sec:ana}, analytic calculations are performed to study the effect of using averaged efficiency in a toy model.
In Sec.~\ref{sec:toy}, we study this effect numerically in toy models assuming net-charge fluctuation and non-uniform acceptance.

\begin{widetext}

\section{Single variable case\label{sec:single}}

Although the main goal of this paper is to derive formulas for the efficiency correction with many efficiency bins, 
in this section we start from the case with a single efficiency bin, because this analysis serves as a simple 
illustration of the multivariate case, which will be addressed in the next section.


\subsection{Cumulants and factorial cumulants}

Let us consider a probability distribution function $P(N)$ for
an integer stochastic variable $N$.
The $m$th order cumulant of $P(N)$ is defined as \cite{Asakawa:2015ybt}
\begin{eqnarray}
\ave{ N^m }_{\rm c} 
= \frac{\partial^m}{\partial\theta^m} K(\theta)\Big|_{\theta=0},
\label{eq:<n^m>c}
\end{eqnarray}
with the cumulant generating function 
\begin{eqnarray}
K(\theta) = \ln \sum_N e^{N\theta} P(N) = \ln \ave{ e^{N\theta} }.
\label{eq:K_c}
\end{eqnarray}
In this study, we fully make use of another set of quantities called 
factorial cumulants $\langle N^m \rangle_{\rm fc}$, which are
defined as 
\begin{eqnarray}
\ave{ N^m }_{\rm fc} 
= \frac{\partial^m}{\partial s^m} K_{\rm f}(s)\Big|_{s=1},
\label{eq:<n^m>fc}
\end{eqnarray}
with the factorial-cumulant generating function 
\begin{eqnarray}
K_{\rm f}(s) = \ln \ave{ s^N }.
\label{eq:K_fc}
\end{eqnarray}

Cumulants can be represented by the sum of factorial cumulants, and vice versa.
To obtain these relations, it is convenient to use the fact that 
the generating functions (\ref{eq:K_c}) and (\ref{eq:K_fc})
are related with each other by the change of variables, $s=e^{\theta}$
or $\theta=\ln s$.
The mutual derivatives of $s$ and $\theta$ are given by
\begin{eqnarray}
	\frac{\partial^{m}}{\partial\theta^{m}}s\bigg|_{s=1} &=& 1, \label{eq:stheta}\\
	\frac{\partial^{m}}{\partial s^{m}}\theta\bigg|_{\theta=0} &=& (-1)^{m-1}(m-1)!. \label{eq:thetas}
\end{eqnarray}
Using Eq.~(\ref{eq:stheta}) the first and second order cumulants are converted into factorial cumulants as
\begin{eqnarray}
	\ave{N}_{\rm c} &=& \frac{\partial}{\partial\theta}K 
        = \frac{\partial s}{\partial\theta}\frac{\partial K_{\rm f}}{\partial s} = \ave{N}_{\rm fc}, 
        \label{eq:ctofc_1}\\
	\ave{N^{2}}_{\rm c} &=& \frac{\partial^{2}}{\partial\theta^{2}}K 
        = \frac{\partial}{\partial\theta}\biggl(\frac{\partial s}{\partial\theta}\frac{\partial K_{\rm f}}{\partial s}\biggr) \nonumber \\ 
	&=& \frac{\partial^{2}s}{\partial\theta^{2}}\frac{\partial K_{\rm f}}{\partial s}+\biggl(\frac{\partial s}{\partial\theta}\biggr)^{2}\frac{\partial^{2}K_{\rm f}}{\partial s^{2}} = \ave{N^{2}}_{\rm fc} + \ave{N}_{\rm fc},
\end{eqnarray}
where it is understood that $\theta=0$ or $s=1$ is substituted.
Repeating the same manipulation, 
we obtain the relation up to sixth order as 
\begin{eqnarray}
	\ave{N^{3}}_{\rm c} &=& \ave{N^{3}}_{\rm fc} + 3\ave{N^{2}}_{\rm fc} + \ave{N}_{\rm fc}, \label{eq:ctofc_3} \\
	\ave{N^{4}}_{\rm c} &=& \ave{N^{4}}_{\rm fc} + 6\ave{N^{3}}_{\rm fc} + 7\ave{N^{2}}_{\rm fc} + \ave{N}_{\rm fc}, \\
	\ave{N^{5}}_{\rm c} &=& \ave{N^{5}}_{\rm fc} + 10\ave{N^{4}}_{\rm fc} + 25\ave{N^{3}}_{\rm fc} + 15\ave{N^{2}}_{\rm fc} + \ave{N}_{\rm fc}, \\
	\ave{N^{6}}_{\rm c} &=& \ave{N^{6}}_{\rm fc} + 15\ave{N^{5}}_{\rm fc} + 65\ave{N^{4}}_{\rm fc} + 90\ave{N^{3}}_{\rm fc} + 31\ave{N^{2}}_{\rm fc} + \ave{N}_{\rm fc}.
        \label{eq:ctofc_6}
\end{eqnarray}
Using Eq.~(\ref{eq:thetas}), factorial cumulants can also be expressed in terms of cumulants as
\begin{eqnarray}
	\ave{N}_{\rm fc} &=& \ave{N}_{\rm c}, 
        \label{eq:fctoc_1}\\
	\ave{N^{2}}_{\rm fc} &=& \ave{N^{2}}_{\rm c} - \ave{N}_{\rm c}, \\
	\ave{N^{3}}_{\rm fc} &=& \ave{N^{3}}_{\rm c} - 3\ave{N^{2}}_{\rm c} + 2\ave{N}_{\rm c}, \\
	\ave{N^{4}}_{\rm fc} &=& \ave{N^{4}}_{\rm c} - 6\ave{N^{3}}_{\rm c} + 11\ave{N^{2}}_{\rm c} - 6\ave{N}_{\rm c}, \\
	\ave{N^{5}}_{\rm fc} &=& \ave{N^{5}}_{\rm c} - 10\ave{N^{4}}_{\rm c} + 35\ave{N^{3}}_{\rm c} - 50\ave{N^{2}}_{\rm c} + 24\ave{N}_{\rm c}, \\
	\ave{N^{6}}_{\rm fc} &=& \ave{N^{6}}_{\rm c} - 15\ave{N^{5}}_{\rm c} + 85\ave{N^{4}}_{\rm c} - 225\ave{N^{3}}_{\rm c} + 274\ave{N^{2}}_{\rm c} - 120\ave{N}_{\rm c} , 
\label{eq:fctoc_6}
\end{eqnarray}
which are summarized in a compact form
\begin{eqnarray}
  \ave{N^m}_{\rm fc} &=& \ave{N(N-1)\cdots(N-m+1)}_{\rm c}.
\end{eqnarray}

\subsection{Binomial model}

Next, we consider the efficiency correction of cumulants in the binomial model~\cite{Asakawa:2015ybt}.
We assume that a multiplicity distribution of a particle number 
$N$ is given by $P(N)$.
We then suppose that individual particles are observed with a probability 
$p$, which is independent for different particles.
We denote the number of observed particles as $n$, and 
the distribution of $n$ as $\tilde{P}(n)$.
Then, $\tilde{P}(n)$ is related to $P(N)$ using the binomial 
distribution function as \cite{Asakawa:2015ybt}
\begin{eqnarray}
	\tilde{P}(n) &=& \sum_{N}P(N)B_{p,N}(n), 
        \label{eq:P=PB}
\end{eqnarray}  
where the binomial distribution and its factorial-cumulant generating 
function are given by 
\begin{eqnarray}
  B_{p,N}(n) &=& \frac{N!}{n!(N-n)!}p^{n}(1-p)^{N-n},
  \\
  \overline{k}_{p,N}(s) &=& {\rm ln}\sum_{n} s^n B_{p,N}(n) 
  = N{\rm ln}(1-p+ps).
  \label{eq:gen_fact_binom}
\end{eqnarray}  
We call Eq.~(\ref{eq:P=PB}) as the binomial model.
From Eq.~(\ref{eq:P=PB}), 
we obtain the factorial-cumulant generating functions of 
$\tilde{P}(n)$ as 
\begin{eqnarray}
	\tilde{K_{\rm f}}(s) &=& {\rm ln}\sum_{n}\tilde{P}(n)s^{n}
        = {\rm ln}\sum_{N}P(N)e^{\overline{k}_{p,N}(s)} = {\rm ln}\sum_{N}P(N)(1-p+ps)^{N}
        = K_{\rm f}(1+p(s-1)),
        \label{eq:K_f=K_f}
\end{eqnarray}
where $K_{\rm f}(s) = {\rm ln}\sum_{N}P(N)s^{N}$ is 
the factorial-cumulant generating function of $P(N)$.
From Eq.~(\ref{eq:K_f=K_f}), one obtains
\begin{eqnarray}
  \frac{\partial^{m}}{\partial s^{m}}\tilde{K}_{\rm f}(s)\bigg|_{s=1}
  = p^{m}\frac{\partial^{m}}{\partial s^{m}}K_{\rm f}(s)\bigg|_{s=1}. 
\label{eq:diff_fact}
\end{eqnarray}
From this result and the definition of the factorial cumulant Eq.~(\ref{eq:<n^m>fc}),
we obtain simple relations between the factorial cumulants of $P(N)$ and 
$\tilde{P}(n)$,
\begin{equation}
	\ave{n^{m}}_{\rm fc} = p^{m}\ave{N^{m}}_{\rm fc}. \label{eq:binom_fact_eff}
\end{equation}
We note that the same relation holds for factorial moments, which is 
used in Ref.~\cite{eff_koch} to derive the formula of efficiency correction.

\subsection{Efficiency correction}

In order to perform the efficiency correction,
we have to represent $\ave{N^{m}}_{\rm c}$ 
in terms of $\ave{n^m}_{\rm c}$.
These relations are obtained by the following three steps:
\begin{enumerate}
\item
Convert $\ave{N^{m}}_{\rm c}$ into $\ave{N^{m}}_{\rm fc}$ using 
Eqs.~(\ref{eq:ctofc_1})--(\ref{eq:ctofc_6}).
\item
Convert $\ave{N^{m}}_{\rm fc}$ into $\ave{n^{m}}_{\rm fc}$ using 
Eq.~(\ref{eq:binom_fact_eff}).
\item
Convert $\ave{n^{m}}_{\rm fc}$ into $\ave{n^{m}}_{\rm c}$ using 
Eqs.~(\ref{eq:fctoc_1})--(\ref{eq:fctoc_6}).
\end{enumerate}
The specific procedures for the first and second orders are as follows:
\begin{eqnarray}
	\ave{N}_{\rm c} &=& \ave{N}_{\rm fc} = \frac{1}{p}\ave{n}_{\rm fc} = \frac{1}{p}\ave{n}_{\rm c}, \label{eq:final_C1} \\
	\ave{N^{2}}_{\rm c} &=& \ave{N^{2}}_{\rm fc} + \ave{N}_{\rm fc} = \frac{1}{p^{2}}\ave{n^{2}}_{\rm fc} + \frac{1}{p}\ave{n}_{\rm fc} \nonumber \\
			    &=& \frac{1}{p^{2}}\bigl(\ave{n^{2}}_{\rm c}-\ave{n}_{\rm c}\bigr) + \frac{1}{p}\ave{n}_{\rm c} \nonumber \\
			    &=& \frac{1}{p^{2}}\ave{n^{2}}_{\rm c} + \biggl(\frac{1}{p}-\frac{1}{p^{2}}\biggr)\ave{n}_{\rm c}. \label{eq:final_C2}
\end{eqnarray}
Similar manipulation up to sixth order are obtained as 
\begin{eqnarray}
  \ave{N^{3}}_{\rm c} 
  &=& 
  \frac{1}{p^{3}} \ave{n^{3}}_{\rm c}
  + \Bigl(-\frac3{p^3}+\frac3{p^2}\Bigr)\ave{n^{2}}_{\rm c}
  + \Bigl(\frac2{p^3}-\frac3{p^2}+\frac1p \Bigr) \ave{n}_{\rm c} ,
  \\
  \ave{N^{4}}_{\rm c} 
  &=& 
  \frac{1}{p^{4}} \ave{n^{4}}_{\rm c}
  + \Bigl( - \frac6{p^{4}} + \frac6{p^3} \Bigr) \ave{n^{3}}_{\rm c}
  + \Bigl( \frac{11}{p^{4}} - \frac{18}{p^3} + \frac7{p^2} \Bigr) \ave{n^2}_{\rm c}
  + \Bigl( - \frac{6}{p^{4}} + \frac{12}{p^3} - \frac7{p^2} + \frac1p \Bigr) \ave{n}_{\rm c} ,
  \\
  \ave{N^{5}}_{\rm c} 
  &=& 
  \frac{1}{p^{5}} \ave{n^{5}}_{\rm c}
  + \Bigl( - \frac{10}{p^{5}} + \frac{10}{p^{4}} \Bigr) \ave{n^{4}}_{\rm c}
  + \Bigl( \frac{35}{p^{5}} - \frac{60}{p^{4}} + \frac{25}{p^{3}} \Bigr) \ave{n^{3}}_{\rm c}
  + \Bigl( -\frac{50}{p^{5}} + \frac{110}{p^{4}} - \frac{75}{p^{3}} + \frac{15}{p^{2}} \Bigr) \ave{n^{2}}_{\rm c}
  \nonumber \\ &&
  + \Bigl( \frac{24}{p^{5}} - \frac{60}{p^{4}} + \frac{50}{p^{3}} - \frac{15}{p^{2}} + \frac1p \Bigr) \ave{n}_{\rm c} ,
  \\
  \ave{N^{6}}_{\rm c} 
  &=& 
  \frac{1}{p^{6}} \ave{n^{6}}_{\rm c}
  + \Bigl( - \frac{15}{p^{6}} + \frac{15}{p^{5}} \Bigr) \ave{n^{5}}_{\rm c}
  + \Bigl( \frac{85}{p^{6}} - \frac{15\cdot10}{p^{5}} + \frac{65}{p^{4}} \Bigr) \ave{n^{4}}_{\rm c}
  + \Bigl( - \frac{225}{p^{6}} + \frac{15\cdot35}{p^{5}} - \frac{65\cdot6}{p^{4}}+ \frac{90}{p^{3}} \Bigr) \ave{n^{3}}_{\rm c}
  \nonumber \\ &&
  + \Bigl( \frac{274}{p^{6}} - \frac{15\cdot50}{p^{5}} + \frac{65\cdot11}{p^{4}} - \frac{90\cdot3}{p^{3}} + \frac{31}{p^2} \Bigr) \ave{n^{2}}_{\rm c}
  + \Bigl( - \frac{120}{p^{6}} + \frac{15\cdot24}{p^{5}} - \frac{65\cdot6}{p^{4}} + \frac{90\cdot2}{p^{3}} - \frac{31}{p^2} + \frac1p \Bigr) \ave{n}_{\rm c} .
\end{eqnarray}
We can extend the manipulation to much higher orders straightforwardly.
These relations are equivalent to those given in Ref.~\cite{Asakawa:2015ybt},
which are obtained based on the cumulant expansion.


\section{Multivariate case\label{sec:multi}}

\subsection{Cumulant and factorial cumulant}

Next, we extend the analysis in the previous section to the multivariate case.
We consider the probability distribution function 
\begin{eqnarray}
	P(\bm{N}) &=& P(N_{1},N_{2},...,N_{M}),
        \label{eq:P(N)mult}
\end{eqnarray}
for $M$ stochastic variables $N_1,\ N_2,\ \cdots,\ N_M$.
Here, $N_i$ with different $i$ represent, for example, particle numbers 
entering detectors which cover different acceptances.
We then consider the cumulants of charges, which is given by 
the linear combination of $N_i$ \cite{psd_Kitazawa},
\begin{eqnarray}
  Q_{(a)} = \sum_{i=1}^M a_i N_i .
\end{eqnarray}
For example, when one considers the net-baryon number, 
$a_i=1$ and $-1$ for baryons and anti-baryons.
For net-electric charge, $a_i$ represents the electric charge of 
particle $i$.

Defining the cumulant generating function of Eq.~(\ref{eq:P(N)mult})
as 
\begin{eqnarray}
  K(\bm{\theta}) 
  = {\rm ln}\Big[ \sum_{\bm{N}}e^{\theta_{1}N_{1}+...+\theta_{M}N_{M}}P(\bm{N}) \Big],
\end{eqnarray}
the $m$th order cumulant of $Q_{(a)}$ is given by
\begin{eqnarray}
  \ave{ Q_{(a)}^m }_{\rm c} = \partial_{(a)}^m K(\bm{\theta})|_{\bm{\theta}=0}
\end{eqnarray}
with 
\begin{equation}
  \partial_{(a)} = \sum_{i=1}^{M}a_{i}\frac{\partial}{\partial \theta_{i}}.
\end{equation}
Similarly, the mixed cumulants are defined by
\begin{eqnarray}
  \ave{ Q_{(a)} Q_{(b)} }_{\rm c} 
  = \partial_{(a)} \partial_{(b)} K(\bm{\theta}) |_{\bm{\theta}=0},
  \quad
  \ave{ Q_{(a)} Q_{(b)} Q_{(c)} }_{\rm c} 
  = \partial_{(a)} \partial_{(b)} \partial_{(c)} K(\bm{\theta}) |_{\bm{\theta}=0},
\end{eqnarray}
and so forth.
The factorial cumulants of $P(\bm{N})$ are defined from 
the generating function \cite{Kitazawa:2017ljq}
\begin{eqnarray}
  K_{\rm f}(\bm{s}) = {\rm ln} \Big[ 
    \sum_{\bm{N}}P(\bm{N})\prod_{i=1}^{M}s_{i}^{N_{i}} \Big]
\end{eqnarray}
as 
\begin{eqnarray}
  \ave{ Q_{(a)}^m }_{\rm fc} = \bar{\partial}_{(a)}^m K_{\rm f}(\bm{s}) |_{\bm{s}=1}, 
  \quad
  \ave{ Q_{(a)} Q_{(b)} }_{\rm fc} 
  = \bar{\partial}_{(a)} \bar{\partial}_{(b)}   K_{\rm f}(\bm{s}) |_{\bm{s}=1},
\end{eqnarray}
and so forth, with 
\begin{equation}
  \overline{\partial}_{(a)} = \sum_{i=1}^{M}a_{i}\frac{\partial}{\partial s_{i}}.
\end{equation}

The relations between cumulants and factorial cumulants can be obtained 
similarly to the previous section.
Using the fact that $K(\bm{\theta})$ and $K_{\rm f}(\bm{s})$ are connected
with each other by the change of variables $s_i=e^{\theta_i}$, we obtain
\begin{eqnarray}
  \ave{ Q_{(a)}}_{\rm c} &=& \partial_{(a)}K 
  = \sum_{i=1}^{M}a_{i}\frac{\partial}{\partial\theta_{i}}K 
  = \sum_{i=1}^{M}a_{i}\frac{\partial s_{i}}{\partial\theta_{i}}\frac{\partial}{\partial s_{i}}K_{\rm f} 
  = \overline{\partial}_{(a)}K_{\rm f}  = \ave{ Q_{(a)}}_{\rm fc},
  \label{eq:multi_ctofc_1} \\
  \ave{ Q_{(a)}Q_{(b)}}_{\rm c} &=& \partial_{(a)}\partial_{(b)}K 
  = \biggl(\sum_{i=1}^{M}a_{i}\frac{\partial}{\partial\theta_{i}}\biggr)\biggl(\sum_{j=1}^{M}b_{j}\frac{\partial}{\partial\theta_{j}}\biggr)K 
  = \sum_{i=j=1}^{M}a_{i}b_{j}\biggl(\frac{\partial s_{i}}{\partial\theta_{i}}\frac{\partial}{\partial s_{i}}\biggr)\biggl(\frac{\partial s_{j}}{\partial\theta_{j}}\frac{\partial}{\partial s_{j}}\biggr)K_{\rm f}  
  \nonumber \\
  &=& \biggl(\sum_{i=1}^{M}a_{i}\frac{\partial s_{i}}{\partial\theta_{i}}\frac{\partial}{\partial s_{i}}\biggr)\biggl(\sum_{j=1}^{M}b_{j}\frac{\partial s_{j}}{\partial\theta_{j}}\frac{\partial}{\partial s_{j}}\biggr)K_{\rm f}  + \sum_{i=1}^{M}a_{i}b_{i}\frac{\partial^{2}s_{i}}{\partial\theta_{i}^{2}}\frac{\partial}{\partial s_{i}}K_{\rm f}  
  \nonumber \\
  &=& \bigl(\overline{\partial}_{(a)}\overline{\partial}_{(b)} + \overline{\partial}_{(ab)}\bigr)K_{\rm f} = \ave{ Q_{(a)}Q_{(b)}}_{\rm fc} + \ave{ Q_{(ab)} }_{\rm fc} ,
 \label{eq:multi_ctofc_2}
\end{eqnarray}
where we assumed $\bm{\theta}=0$ or $\bm{s}=1$, and 
in the last line we defined 
\begin{eqnarray}
  \overline{\partial}_{(ab)} 
  = \sum_{i=1}^M a_i b_i \frac{\partial}{\partial s_{i}} ,
  \quad
  Q_{(ab)} = \sum_{i=1}^M a_i b_i N_i .
\end{eqnarray}
We also define $\partial_{(ab)}$ and the symbols with more than 
two subscripts, such as $Q_{(abc)}$, in a similar manner.
This manipulation can be extended straightforwardly to arbitrary higher orders.
In this analysis, we use the following relations between $\theta$ and 
$s$ derivatives valid for $\bm{s}=1$:
\begin{eqnarray}
	\partial_{(a)} &=& \overline{\partial}_{(a)},
        \nonumber \\
	\partial_{(a)}\partial_{(b)} &=& \overline{\partial}_{(a)}\overline{\partial}_{(b)} + \overline{\partial}_{(ab)} 
        \nonumber \\ 
	\partial_{(a)}\partial_{(b)}\partial_{(c)} &=& \overline{\partial}_{(a)}\overline{\partial}_{(b)}\overline{\partial}_{(c)} + \overline{\partial}_{(a)}\overline{\partial}_{(bc)} + \overline{\partial}_{(b)}\overline{\partial}_{(ca)} \nonumber \\ 
	&& + \overline{\partial}_{(c)}\overline{\partial}_{(ab)} + \overline{\partial}_{(abc)}, \label{eq:multi_ctofc_3}\nonumber \\  \\
	\partial_{(a)}\partial_{(b)}\partial_{(c)}\partial_{(d)} &=& \overline{\partial}_{(a)}\overline{\partial}_{(b)}\overline{\partial}_{(c)}\overline{\partial}_{(d)} \nonumber \\
	&&  + \Bigl[ \overline{\partial}_{(a)}\overline{\partial}_{(b)}\overline{\partial}_{(cd)} + (6\;{\rm comb.}) \Bigr]  \nonumber \\
	&&  + \Bigl[ \overline{\partial}_{(a)}\overline{\partial}_{(bcd)} + (4\;{\rm comb.}) \Bigr] \nonumber  \\
	&&  + \Bigl[ \overline{\partial}_{(ab)}\overline{\partial}_{(cd)} + (3\;{\rm comb.}) \Bigr] \nonumber  \\
	&&  + \overline{\partial}_{(abcd)}, \\
	\partial_{(a)}\partial_{(b)}\partial_{(c)}\partial_{(d)}\partial_{(e)} &=& \overline{\partial}_{(a)}\overline{\partial}_{(b)}\overline{\partial}_{(c)}\overline{\partial}_{(d)}\overline{\partial}_{(e)} \nonumber \\
	&&  + \Bigl[ \overline{\partial}_{(a)}\overline{\partial}_{(b)}\overline{\partial}_{(c)}\overline{\partial}_{(de)} + (10\;{\rm comb.}) \Bigr] \nonumber \\
	&&  + \Bigl[ \overline{\partial}_{(a)}\overline{\partial}_{(b)}\overline{\partial}_{(cde)} + (10\;{\rm comb.}) \Bigr] \nonumber \\
	&&  + \Bigl[ \overline{\partial}_{(a)}\overline{\partial}_{(bc)}\overline{\partial}_{(de)} + (15\;{\rm comb.}) \Bigr] \nonumber \\
	&&  + \Bigl[ \overline{\partial}_{(a)}\overline{\partial}_{(bcde)} + (5\;{\rm comb.})  \Bigr] \nonumber  \\
	&&  + \Bigl[ \overline{\partial}_{(ab)}\overline{\partial}_{(cde)} + (10\;{\rm comb.}) \Bigr] \nonumber \\
	&&  +  \overline{\partial}_{(abcde)}, \\
	\partial_{(a)}\partial_{(b)}\partial_{(c)}\partial_{(d)}\partial_{(e)}\partial_{(f)} &=& \overline{\partial}_{(a)}\overline{\partial}_{(b)}\overline{\partial}_{(c)}\overline{\partial}_{(d)}\overline{\partial}_{(e)}\overline{\partial}_{(f)} \nonumber \\
	&&  + \Bigl[ \overline{\partial}_{(a)}\overline{\partial}_{(b)}\overline{\partial}_{(c)}\overline{\partial}_{(d)}\overline{\partial}_{(ef)} + (15\;{\rm comb.}) \Bigr] \nonumber \\
	&&  + \Bigl[ \overline{\partial}_{(a)}\overline{\partial}_{(b)}\overline{\partial}_{(c)}\overline{\partial}_{(def)} + (20\;{\rm comb.}) \Bigr] \nonumber \\
	&&  + \Bigl[ \overline{\partial}_{(a)}\overline{\partial}_{(b)}\overline{\partial}_{(cd)}\overline{\partial}_{(ef)} + (45\;{\rm comb.}) \Bigr] \nonumber \\
	&&  + \Bigl[ \overline{\partial}_{(a)}\overline{\partial}_{(b)}\overline{\partial}_{(cdef)} + (15\;{\rm comb.}) \Bigr] \nonumber \\
	&&  + \Bigl[ \overline{\partial}_{(a)}\overline{\partial}_{(bc)}\overline{\partial}_{(def)} + (60\;{\rm comb.}) \Bigr] \nonumber \\
	&&  + \Bigl[ \overline{\partial}_{(ab)}\overline{\partial}_{(cd)}\overline{\partial}_{(ef)} + (15\;{\rm comb.}) \Bigr] \nonumber \\
	&&  + \Bigl[ \overline{\partial}_{(abc)}\overline{\partial}_{(def)} + (10\;{\rm comb.}) \Bigr] \nonumber \\
	&&  + \Bigl[ \overline{\partial}_{(ab)}\overline{\partial}_{(cdef)} + (15\;{\rm comb.}) \Bigr] \nonumber \\
	&&  + \Bigl[ \overline{\partial}_{(a)}\overline{\partial}_{(bcdef)} + (6\;{\rm comb.}) \Bigr]  \nonumber \\
	&&  + \overline{\partial}_{(abcdef)}.
\end{eqnarray}
Here, $(\rm comb.)$ represents terms obtained by all possible combinations of subscripts, for example,
\begin{eqnarray}
  \overline{\partial}_{(a)}\overline{\partial}_{(b)}\overline{\partial}_{(cd)} + (6\;{\rm comb.}) 
  &=& \overline{\partial}_{(a)}\overline{\partial}_{(b)}\overline{\partial}_{(cd)}
  + \overline{\partial}_{(a)}\overline{\partial}_{(c)}\overline{\partial}_{(bd)}
  + \overline{\partial}_{(a)}\overline{\partial}_{(d)}\overline{\partial}_{(bc)}
  \nonumber \\
  &&+ \overline{\partial}_{(b)}\overline{\partial}_{(c)}\overline{\partial}_{(ad)}
  + \overline{\partial}_{(b)}\overline{\partial}_{(d)}\overline{\partial}_{(ac)}
  + \overline{\partial}_{(c)}\overline{\partial}_{(d)}\overline{\partial}_{(ab)}.
\end{eqnarray}
Note that the number shows the total number of the combinations.
The conversions from factorial cumulants to cumulants can be carried 
out with the following relations valid for $\bm{\theta}=0$:
\begin{eqnarray}
	\overline{\partial}_{(a)} &=& \partial_{(a)}, \label{eq:multi_fctoc_1} \\
	\overline{\partial}_{(a)}\overline{\partial}_{(b)} &=& \partial_{(a)}\partial_{(b)} - \partial_{(ab)}, \label{eq:multi_fctoc_2}\\
	\overline{\partial}_{(a)}\overline{\partial}_{(b)}\overline{\partial}_{(c)} &=& \partial_{(a)}\partial_{(b)}\partial_{(c)} - \partial_{(a)}\partial_{(bc)} - \partial_{(b)}\partial_{(ca)} \nonumber \\
	&& - \partial_{(c)}\partial_{(ab)} + 2 \partial_{(abc)}. \label{eq:multi_fctoc_3}\\
	\overline{\partial}_{(a)}\overline{\partial}_{(b)}\overline{\partial}_{(c)}\overline{\partial}_{(d)} &=& \partial_{(a)}\partial_{(b)}\partial_{(c)}\partial_{(d)} \nonumber \\
	 &&- \Bigl[ \partial_{(a)}\partial_{(b)}\partial_{(cd)} + (6\;{\rm comb.}) \Bigr] \nonumber \\
	 &&+2\Bigl[ \partial_{(a)}\partial_{(bcd)} + (4\;{\rm comb.}) \Bigr] \nonumber \\
	 &&+ \Bigl[ \partial_{(ab)}\partial_{(cd)} + (3\;{\rm comb.}) \Bigr] \nonumber \\
	 &&-6\partial_{(abcd)}, \\
	\overline{\partial}_{(a)}\overline{\partial}_{(b)}\overline{\partial}_{(c)}\overline{\partial}_{(d)}\overline{\partial}_{(e)} &=& \partial_{(a)}\partial_{(b)}\partial_{(c)}\partial_{(d)}\partial_{(e)} \nonumber \\
	&& -   \Bigl[ \partial_{(a)}\partial_{(b)}\partial_{(c)}\partial_{(de)} + (10\;{\rm comb.}) \Bigr] \nonumber \\
	&& +2  \Bigl[ \partial_{(a)}\partial_{(b)}\partial_{(cde)} + (10\;{\rm comb.}) \Bigr] \nonumber \\
	&& +   \Bigl[ \partial_{(a)}\partial_{(bc)}\partial_{(de)} + (15\;{\rm comb.}) \Bigr] \nonumber \\
	&& -6  \Bigl[ \partial_{(a)}\partial_{(bcde)} + (5\;{\rm comb.})  \Bigr] \nonumber \\
	&& -2  \Bigl[ \partial_{(ab)}\partial_{(cde)} + (10\;{\rm comb.}) \Bigr] \nonumber \\
	&& +24 \partial_{(abcde)}, \\
	\overline{\partial}_{(a)}\overline{\partial}_{(b)}\overline{\partial}_{(c)}\overline{\partial}_{(d)}\overline{\partial}_{(e)}\overline{\partial}_{(f)} &=& \partial_{(a)}\partial_{(b)}\partial_{(c)}\partial_{(d)}\partial_{(e)}\partial_{(f)} \nonumber \\
	&&  -   \Bigl[ \partial_{(a)}\partial_{(b)}\partial_{(c)}\partial_{(d)}\partial_{(ef)} + (15\;{\rm comb.}) \Bigr] \nonumber \\
	&&  + 2 \Bigl[ \partial_{(a)}\partial_{(b)}\partial_{(c)}\partial_{(def)} + (20\;{\rm comb.}) \Bigr] \nonumber \\
	&&  + \Bigl[ \partial_{(a)}\partial_{(b)}\partial_{(cd)}\partial_{(ef)} + (45\;{\rm comb.}) \Bigr] \nonumber \\
	&&  - 6 \Bigl[ \partial_{(a)}\partial_{(b)}\partial_{(cdef)} + (15\;{\rm comb.}) \Bigr] \nonumber \\
	&&  - 2 \Bigl[ \partial_{(a)}\partial_{(bc)}\partial_{(def)} + (60\;{\rm comb.}) \Bigr] \nonumber \\
	&&  -   \Bigl[ \partial_{(ab)}\partial_{(cd)}\partial_{(ef)} + (15\;{\rm comb.}) \Bigr] \nonumber \\
	&&  + 4 \Bigl[ \partial_{(abc)}\partial_{(def)} + (10\;{\rm comb.}) \Bigr] \nonumber \\
	&&  + 6 \Bigl[ \partial_{(ab)}\partial_{(cdef)} + (15\;{\rm comb.}) \Bigr] \nonumber \\
	&&  + 24\Bigl[ \partial_{(a)}\partial_{(bcdef)} + (6\;{\rm comb.})  \Bigr] \nonumber \\
	&&  - 120\partial_{(abcdef)}.
\end{eqnarray}

\subsection{Efficiency correction in binomial model}

Next, we extend the binomial model Eq.~(\ref{eq:P=PB}) to 
the multivariate case.
We suppose that a particle labeled by $i$ 
is observed with efficiency $p_{i}$.
Assuming the independence of the efficiencies of individual particles,
the probability distribution function $\tilde{P}(\bm{n})$ of observed 
particle numbers $n_i$ is related to $P(\bm{N})$ as \cite{psd_Kitazawa}
\begin{equation}
	\tilde{P}(\bm{n}) = \sum_{\bm{N}}P(\bm{N})\prod_{i=1}^{M}B_{p_{i},N_{i}}(n_{i}).
\end{equation}
The factorial-cumulant generating function of $\tilde{P}(\bm{n})$ 
is then given by 
\begin{eqnarray}
  \tilde{K}_{\rm f}(\bm{s}) 
  = {\rm ln}\sum_{\bm{N}}P(\bm{N})\prod_{i=1}^{M}(1+p_{i}(s_{i}-1))^{N_{i}}
  = {K}_{\rm f}(\bm{s}') ,
  \label{eq:multi_binom_fact_eff}
\end{eqnarray}
with $s'_i = 1 + p_i(s_i-1)$.
From Eq.~(\ref{eq:multi_binom_fact_eff}), one finds that 
$\overline{\partial}_{(a)}\tilde{K}_{\rm f} = \overline{\partial}_{(ap)}K_{\rm f}$
and 
\begin{eqnarray}
  \overline{\partial}_{(a)}K_{\rm f} = \overline{\partial}_{(a/p)}\tilde{K}_{\rm f} ,
  \quad
  \label{eq:multi_fact_eff}
  \overline{\partial}_{(a)} \overline{\partial}_{(b)} K_{\rm f} 
  = \overline{\partial}_{(a/p)}\overline{\partial}_{(b/p)}\tilde{K}_{\rm f} ,
\end{eqnarray}
and so forth, where it is understood that $\bm{s}=1$ is substituted
and $\overline{\partial}_{(a/p)}=\sum_{i=1}^M (a_i/p_i) (\partial/\partial s_i)$.
Equation~(\ref{eq:multi_fact_eff}) connects the factorial cumulants of 
$\tilde{P}(\bm{n})$ and $P(\bm{n})$.

For the efficiency correction, one must represent the cumulants of 
$P(\bm{n})$ by those of $\tilde{P}(\bm{n})$.
Similar to the procedure in Sec.~\ref{sec:single}, 
these relations are obtained by the following steps:
\begin{enumerate}
\item
Convert a cumulant of $P(\bm{N})$ into factorial cumulants.
\item
Convert the factorial cumulants of $P(\bm{N})$ 
into factorial cumulants of $\tilde{P}(\bm{N})$.
\item
Convert the factorial cumulants of $\tilde{P}(\bm{N})$
into cumulants.
\end{enumerate}
The explicit manipulation up to the third order is shown as follows:
\begin{eqnarray}
	\ave{Q_{(a)}}_{\rm c} &=& \ave{Q_{(a)}}_{\rm fc} = \ave{q_{(a/p)}}_{\rm fc} = \ave{q_{(a/p)}}_{\rm c}, \\
	\ave{Q_{(a)}^{2}}_{\rm c} &=& \ave{Q_{(a)}^{2}}_{\rm fc} + \ave{Q_{(a^{2})}}_{\rm fc} = \ave{q_{(a/p)}^{2}}_{\rm fc} + \ave{q_{(a^{2}/p)}}_{\rm fc} \nonumber \\
				  &=& \ave{q_{(a/p)}^{2}}_{\rm c} - \ave{q_{(a^{2}/p^{2})}}_{\rm c} + \ave{q_{(a^{2}/p)}}_{\rm c}, \\
	\ave{Q_{(a)}^{3}}_{\rm c} &=& \ave{Q_{(a)}^{3}}_{\rm fc} + 3\ave{Q_{(a)}Q_{(a^{2})}}_{\rm fc} + \ave{Q_{(a^{3})}}_{\rm fc} = \ave{q_{(a/p)}^{3}}_{\rm fc} + 3\ave{q_{(a/p)}q_{(a^{2}/p)}}_{\rm fc} + \ave{q_{(a^{3}/p)}}_{\rm fc} \nonumber \\
				  &=& \ave{q_{(a/p)}^{3}}_{\rm c} - 3\ave{q_{(a/p)}q_{(a^{2}/p^{2})}}_{\rm c} + 2\ave{q_{(a^{3}/p^{3})}}_{\rm c} + 3\bigl(\ave{q_{(a/p)}q_{(a^{2}/p)}}_{\rm c}-\ave{q_{(a^{3}/p^{2})}}_{\rm c}\bigr) + \ave{q_{(a^{3}/p)}} ,
\end{eqnarray}
where we defined the linear combination of $n_i$ as
\begin{equation}
  q_{(a)} = \sum_{i=1}^{M}a_{i}n_{i} , 
  \quad
  q_{(ab/p)} \equiv \sum_{i=1}^{M}(a_{i} b_i/p_i) n_{i},
\end{equation}
and so forth.
The explicit results up to the sixth order are given by 
\begin{eqnarray}
	\bigl<Q\bigr>_{\rm c} &=& \ave{q_{(1,1)}}_{\rm c}, 
        \label{eq:mk_1} \\ \nonumber \\
	\bigl<Q^{2}\bigr>_{\rm c} &=& \ave{q_{(1,1)}^{2}}_{\rm c} + \ave{q_{(2,1)}}_{\rm c} - \ave{q_{(2,2)}}_{\rm c}, 
        \label{eq:mk_2} \\ \nonumber \\
	\bigl<Q^{3}\bigr>_{\rm c} 
        &=& \ave{q_{(1,1)}^{3}}_{\rm c} 
        + 3\ave{q_{(1,1)}q_{(2,1)}}_{\rm c} - 3\ave{q_{(1,1)}q_{(2,2)}}_{\rm c}
        + \ave{q_{(3,1)}}_{\rm c} - 3\ave{q_{(3,2)}}_{\rm c}
        + 2\ave{q_{(3,3)}}_{\rm c}, \label{eq:mk_3}
        \\ \nonumber \\
	\bigl<Q^{4}\bigr>_{\rm c} 
        &=& \ave{q_{(1,1)}^{4}}_{\rm c} 
        + 6\ave{q_{(1,1)}^{2}q_{(2,1)}}_{\rm c} - 6\ave{q_{(1,1)}^{2}q_{(2,2)}}_{\rm c} + 4\ave{q_{(1,1)}q_{(3,1)}}_{\rm c} + 3\ave{q_{(2,1)}^{2}}_{\rm c}
	\nonumber \\
	&& + 3\ave{q_{(2,2)}^{2}}_{\rm c} - 12\ave{q_{(1,1)}q_{(3,2)}}_{\rm c} + 8\ave{q_{(1,1)}q_{(3,3)}}_{\rm c} -  6\ave{q_{(2,1)}q_{(2,2)}}_{\rm c}
        \nonumber \\
	&& + \ave{q_{(4,1)}}_{\rm c} - 7\ave{q_{(4,2)}}_{\rm c} 
        + 12\ave{q_{(4,3)}}_{\rm c} - 6\ave{q_{(4,4)}}_{\rm c}, 
	\label{eq:mk_4} \\ \nonumber \\
	\bigl<Q^{5}\bigr>_{\rm c} &=& \ave{q_{(1,1)}^{5}}_{\rm c} 
				+ 10\ave{q_{(1,1)}^{3}q_{(2,1)}}_{\rm c} - 10\ave{q_{(1,1)}^{3}q_{(2,2)}}_{\rm c} + 10\ave{q_{(1,1)}^{2}q_{(3,1)}}_{\rm c} - 30\ave{q_{(1,1)}^{2}q_{(3,2)}}_{\rm c}  \nonumber \\
			     && + 20\ave{q_{(1,1)}^{2}q_{(3,3)}}_{\rm c} + 15\ave{q_{(2,2)}^{2}q_{(1,1)}}_{\rm c} + 15\ave{q_{(2,1)}^{2}q_{(1,1)}}_{\rm c} - 30\ave{q_{(1,1)}q_{(2,1)}q_{(2,2)}}_{\rm c} \nonumber \\ 
			     && +  5\ave{q_{(1,1)}q_{(4,1)}}_{\rm c} - 35\ave{q_{(1,1)}q_{(4,2)}}_{\rm c} + 60\ave{q_{(1,1)}q_{(4,3)}}_{\rm c} - 30\ave{q_{(1,1)}q_{(4,4)}}_{\rm c}  \nonumber \\
			     && + 10\ave{q_{(2,1)}q_{(3,1)}}_{\rm c} - 30\ave{q_{(2,1)}q_{(3,2)}}_{\rm c} + 20\ave{q_{(2,1)}q_{(3,3)}}_{\rm c}    \nonumber \\ 
			     && - 10\ave{q_{(2,2)}q_{(3,1)}}_{\rm c} + 30\ave{q_{(2,2)}q_{(3,2)}}_{\rm c} - 20\ave{q_{(2,2)}q_{(3,3)}}_{\rm c}   \nonumber \\
			     && + \ave{q_{(5,1)}}_{\rm c} - 15\ave{q_{(5,2)}}_{\rm c} + 50\ave{q_{(5,3)}}_{\rm c} - 60\ave{q_{(5,4)}}_{\rm c} + 24\ave{q_{(5,5)}}_{\rm c}, \label{eq:mk_5} \\ \nonumber \\
	\bigl<Q^{6}\bigr>_{\rm c}&=& \ave{q_{(1,1)}^{6}}_{\rm c} + 15\ave{q_{(1,1)}^{4}q_{(2,1)}}_{\rm c} - 15\ave{q_{(1,1)}^{4}q_{(2,2)}}_{\rm c} + 20\ave{q_{(1,1)}^{3}q_{(3,1)}}_{\rm c} - 60\ave{q_{(1,1)}^{3}q_{(3,2)}}_{\rm c}  \nonumber \\ 
			     && + 40\ave{q_{(1,1)}^{3}q_{(3,3)}}_{\rm c} - 90\ave{q_{(1,1)}^{2}q_{(2,2)}q_{(2,1)}}_{\rm c} + 45\ave{q_{(1,1)}^{2}q_{(2,1)}^{2}}_{\rm c} + 45\ave{q_{(1,1)}^{2}q_{(2,2)}^{2}}_{\rm c} \nonumber \\ 
			     && + 15\ave{q_{(2,1)}^{3}}_{\rm c} - 15\ave{q_{(2,2)}^{3}}_{\rm c} + 15\ave{q_{(1,1)}^{2}q_{(4,1)}}_{\rm c} - 105\ave{q_{(1,1)}^{2}q_{(4,2)}}_{\rm c} + 180\ave{q_{(1,1)}^{2}q_{(4,3)}}_{\rm c} - 90\ave{q_{(1,1)}^{2}q_{(4,4)}}_{\rm c} \nonumber \\ 
			     && - 45\ave{q_{(2,1)}^{2}q_{(2,2)}}_{\rm c} + 45\ave{q_{(2,2)}^{2}q_{(2,1)}}_{\rm c} + 60\ave{q_{(1,1)}q_{(2,1)}q_{(3,1)}}_{\rm c} - 180\ave{q_{(1,1)}q_{(2,1)}q_{(3,2)}}_{\rm c}    \nonumber \\ 
			     && + 120\ave{q_{(1,1)}q_{(2,1)}q_{(3,3)}}_{\rm c} - 60\ave{q_{(1,1)}q_{(2,2)}q_{(3,1)}}_{\rm c} + 180\ave{q_{(1,1)}q_{(2,2)}q_{(3,2)}}_{\rm c} - 120\ave{q_{(1,1)}q_{(2,2)}q_{(3,3)}}_{\rm c}  \nonumber \\
			     && + 6\ave{q_{(1,1)}q_{(5,1)}}_{\rm c} - 90\ave{q_{(1,1)}q_{(5,2)}}_{\rm c} + 300\ave{q_{(1,1)}q_{(5,3)}}_{\rm c} - 360\ave{q_{(1,1)}q_{(5,4)}}_{\rm c} + 144\ave{q_{(1,1)}q_{(5,5)}}_{\rm c} \nonumber \\
			     && + 15\ave{q_{(2,1)}q_{(4,1)}}_{\rm c} - 105\ave{q_{(2,1)}q_{(4,2)}}_{\rm c} + 180\ave{q_{(2,1)}q_{(4,3)}}_{\rm c} - 90\ave{q_{(2,1)}q_{(4,4)}}_{\rm c}   \nonumber \\
			     && - 15\ave{q_{(2,2)}q_{(4,1)}}_{\rm c} + 105\ave{q_{(2,2)}q_{(4,2)}}_{\rm c} - 180\ave{q_{(2,2)}q_{(4,3)}}_{\rm c} + 90\ave{q_{(2,2)}q_{(4,4)}}_{\rm c} \nonumber \\
			     && + 10\ave{q_{(3,1)}^{2}}_{\rm c} - 60\ave{q_{(3,1)}q_{(3,2)}}_{\rm c}+ 40\ave{q_{(3,1)}q_{(3,3)}}_{\rm c} + 90\ave{q_{(3,2)}^{2}}_{\rm c} - 120\ave{q_{(3,2)}q_{(3,3)}}_{\rm c} + 40\ave{q_{(3,3)}^{2}}_{\rm c}      \nonumber \\ 
			     && + \ave{q_{(6,1)}}_{\rm c} - 31\ave{q_{(6,2)}}_{\rm c} + 180\ave{q_{(6,3)}}_{\rm c} - 390\ave{q_{(6,4)}}_{\rm c} + 360\ave{q_{(6,5)}}_{\rm c}  - 120\ave{q_{(6,6)}}_{\rm c},
        \label{eq:mk_6} \nonumber \\ 
\end{eqnarray}
where we used 
\begin{equation}
	q_{(r,s)} = q_{(a^r/p^s)}
        = \sum_{i=1}^{M} (a_{i}^{r}/p_{i}^{s}) n_{i}. 
        \label{eq:multi_q}
\end{equation}
In Appendix~\ref{sec:net}, we show a specific example of these results
for the net-particle number with $M=2$.

In Eqs.~(\ref{eq:mk_1})--(\ref{eq:mk_6}),
the cumulants of $P(\bm{N})$ are expressed in terms of the (mixed) 
cumulants of $\tilde{P}(\bm{n})$.
These formulas thus can be used for the efficiency correction.
We note that the number of cumulants does not depends on the number 
of efficiency bins $M$.
This property is contrasted to the method proposed in 
Refs.~\cite{eff_psd_volker,eff_xiaofeng}, in which the number of 
expectation values to be calculated increases as $\sim M^m$ 
for $m$th order cumulant.
The numerical cost for the efficiency correction with 
Eqs.~(\ref{eq:mk_1})--(\ref{eq:mk_6}) thus is drastically reduced 
compared to the formulas proposed in Refs.~\cite{eff_psd_volker,eff_xiaofeng}
for large $M$.
In the formulas proposed in Ref.~\cite{psd_Kitazawa},
the number of terms is much more reduced compared to 
Eqs.~(\ref{eq:mk_1})--(\ref{eq:mk_6})
and thus the numerical cost is smaller than our method.
However, the derivation in Ref.~\cite{psd_Kitazawa} is complicated 
and it is quite difficult to extend the analysis 
in Ref.~\cite{psd_Kitazawa} to sixth and much higher orders.
We have numerically verified that our method gives completely the same 
result as those in Refs.~\cite{eff_psd_volker,eff_xiaofeng} and Ref.~\cite{psd_Kitazawa}. 
In actual analyses, it would be convenient to implement the derivation 
of Eqs.~(\ref{eq:mk_1})--(\ref{eq:mk_6}) as a numerical algorithm.

\subsection{Mixed cumulants}

So far, we considered the efficiency correction of the cumulants of 
a single charge $Q_{(a)}$.
Exactly the same discussion can be applied to the efficiency correction of 
mixed cumulants, which probe correlations between different conserved 
quantities, e.g, net-baryon, net-strangeness, and net-charge.
Below we show the formulas for mixed cumulants 
$\bigl<Q_{(x)}^{m_1}Q_{(y)}^{m_2}\bigr>$ up to fourth order ($m_1+m_2\leqq4$):
\begin{eqnarray}
  \ave{ Q_{(x)}Q_{(y)} }_{\rm c}
  &=& \ave{q_{(1,0,1)}q_{(0,1,1)}}_{\rm c} + \ave{q_{(1,1,1)}}_{\rm c} - \ave{q_{(1,1,2)}}_{\rm c},
  \\ 
  \nonumber \\ 
  \ave{ Q_{(x)}^2Q_{(y)} }_{\rm c}
	&=& \ave{q_{(1,0,1)}^{2}q_{(0,1,1)}}_{\rm c} 
	+ 2\ave{q_{(1,0,1)}q_{(1,1,1)}}_{\rm c} 
	- 2\ave{q_{(1,0,1)}q_{(1,1,2)}}_{\rm c} 
	+ \ave{q_{(0,1,1)}q_{(2,0,1)}}_{\rm c} 
	- \ave{q_{(0,1,1)}q_{(2,0,2)}}_{\rm c} 
	\nonumber \\ &&
	+ \ave{q_{(2,1,1)}}_{\rm c} -3\ave{q_{(2,1,2)}}_{\rm c} + \ave{q_{(2,1,3)}}_{\rm c},
  \\ 
  \nonumber \\ 
  \ave{ Q_{(x)}^2Q_{(y)}^2 }_{\rm c}
	&=& \ave{q_{(1,0,1)}^{2}q_{(0,1,1)}^{2}}_{\rm c}
	\nonumber \\ &&
	+ \ave{q_{(1,0,1)}^{2}q_{(0,2,1)}}_{\rm c}
	- \ave{q_{(1,0,1)}^{2}q_{(0,2,2)}}_{\rm c}
	+ \ave{q_{(0,1,1)}^{2}q_{(2,0,1)}}_{\rm c}
	- \ave{q_{(0,1,1)}^{2}q_{(2,0,2)}}_{\rm c}
	\nonumber \\ &&
	+4\ave{q_{(1,0,1)}q_{(0,1,1)}q_{(1,1,1)}}_{\rm c}
	-4\ave{q_{(1,0,1)}q_{(0,1,1)}q_{(1,1,2)}}_{\rm c}
	\nonumber \\ &&
	+2\ave{q_{(1,0,1)}q_{(1,2,1)}}_{\rm c}
	-6\ave{q_{(1,0,1)}q_{(1,2,2)}}_{\rm c}
	+4\ave{q_{(1,0,1)}q_{(1,2,3)}}_{\rm c}
	\nonumber \\ &&
	+2\ave{q_{(0,1,1)}q_{(2,1,1)}}_{\rm c}
	-6\ave{q_{(0,1,1)}q_{(2,1,2)}}_{\rm c}
	+4\ave{q_{(0,1,1)}q_{(2,1,3)}}_{\rm c}
	\nonumber \\ &&
	-4\ave{q_{(1,1,1)}q_{(1,1,2)}}_{\rm c}
	+2\ave{q_{(1,1,1)}^{2}}_{\rm c}
	+2\ave{q_{(1,1,2)}^{2}}_{\rm c}
	\nonumber \\ &&
	+ \ave{q_{(2,0,1)}q_{(0,2,1)}}_{\rm c}
	- \ave{q_{(2,0,1)}q_{(0,2,2)}}_{\rm c}
	- \ave{q_{(2,0,2)}q_{(0,2,1)}}_{\rm c}
	+ \ave{q_{(2,0,2)}q_{(0,2,2)}}_{\rm c}
	\nonumber \\ &&
	+ \ave{q_{(2,2,1)}}_{\rm c}
	-7\ave{q_{(2,2,2)}}_{\rm c}
	+12\ave{q_{(2,2,3)}}_{\rm c}
	-6\ave{q_{(2,2,4)}}_{\rm c},
  \\ 
  \nonumber \\ 
  \ave{ Q_{(x)}^3 Q_{(y)} }_{\rm c}
  &=& \ave{q_{(1,0,1)}^{3}q_{(0,1,1)}}_{\rm c}
	\nonumber \\ &&
	+3\ave{q_{(1,0,1)}^{2}q_{(1,1,1)}}_{\rm c}
	-3\ave{q_{(1,0,1)}^{2}q_{(1,1,2)}}_{\rm c}
	+3\ave{q_{(2,0,1)}q_{(1,0,1)}q_{(0,1,1)}}_{\rm c}
	-3\ave{q_{(2,0,2)}q_{(1,0,1)}q_{(0,1,1)}}_{\rm c}
	\nonumber \\ &&
	+3\ave{q_{(1,0,1)}q_{(2,1,1)}}_{\rm c}
	-9\ave{q_{(1,0,1)}q_{(2,1,2)}}_{\rm c}
	+6\ave{q_{(1,0,1)}q_{(2,1,3)}}_{\rm c}
	\nonumber \\ &&
	+3\ave{q_{(2,0,1)}q_{(1,1,1)}}_{\rm c}
	-3\ave{q_{(2,0,1)}q_{(1,1,2)}}_{\rm c}
	-3\ave{q_{(2,0,2)}q_{(1,1,1)}}_{\rm c}
	+3\ave{q_{(2,0,2)}q_{(1,1,2)}}_{\rm c}
	\nonumber \\ &&
	+ \ave{q_{(3,0,1)}q_{(0,1,1)}}_{\rm c}
	-3\ave{q_{(3,0,2)}q_{(0,1,1)}}_{\rm c}
	+2\ave{q_{(3,0,3)}q_{(0,1,1)}}_{\rm c}
	\nonumber \\ &&
	+ \ave{q_{(3,1,1)}}_{\rm c}
	-7\ave{q_{(3,1,2)}}_{\rm c}
	+12\ave{q_{(3,1,3)}}_{\rm c}
	-6\ave{q_{(3,1,4)}}_{\rm c},
\end{eqnarray}
where we used the symbol
\begin{equation}
	q_{(r,s,t)} = q_{(x^{r}y^{s}/p^t)}
        = \sum_{i=1}^{M} (x_{i}^{r}y_{i}^{s}/p_{i}^{t}) n_{i}. 
        \label{eq:multi_qmix}
\end{equation}

\end{widetext}
\subsection{Calculation cost}

Finally we discuss how efficient Eqs.~(\ref{eq:mk_1})--(\ref{eq:mk_6}) are 
compared to the conventional method based on factorial moments~\cite{eff_psd_volker}.
In the conventional method,
a cumulant is decomposed into mixed factorial moments.
In this decomposition for an $m$th order cumulant,
all possible combinations of mixed factorial moments 
between different efficiency bins with order $r$
satisfying $r\le m$ appear.
The number of the $r$th order factorial moments is given by 
${}_{r+M-1}C_{r}$ with $M$ being different efficiency bins.
The total number of the mixed factorial moments satisfying $r\le m$ is 
thus given by 
\begin{equation}
	N_{m}^{\rm fm} = \sum_{r=1}^{m}{}_{r+M-1}C_{r} = _{m+M}C_{m}-1.
	\label{eq:cost_fm}
\end{equation}
When this method is adopted to the analysis of the net-particle number,
the numbers of efficiency bins of particle and anti-particle are
given by $M/2$.

\begin{figure}[H]
	\begin{center}
	\includegraphics[width=70mm]{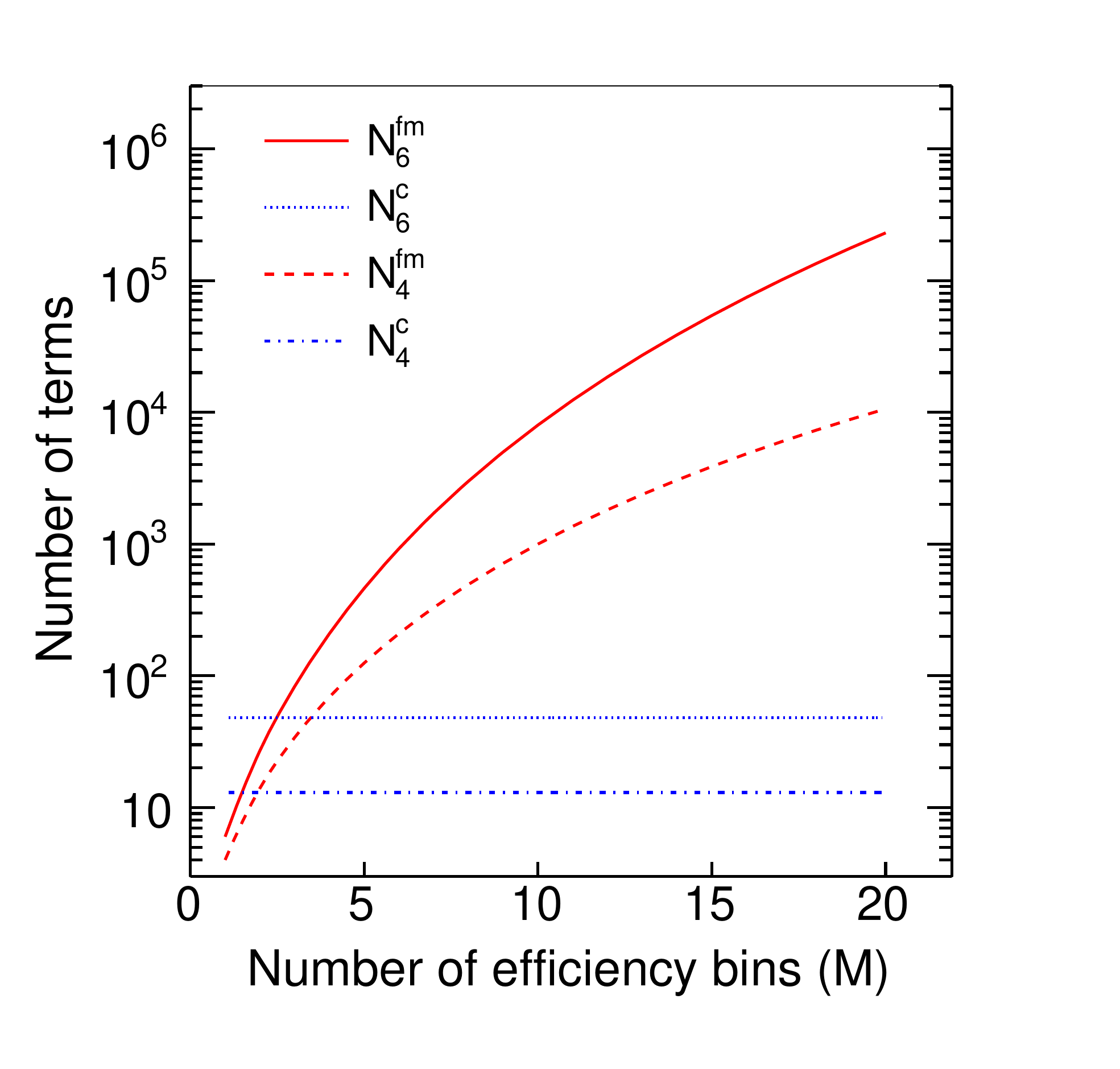}
	\end{center}
	\caption{
          Number of terms (factorial moments or cumulants) to be calculated
	  in the conventional method based on the factorial moments~\cite{eff_psd_volker} and 
          the new method in this paper, Eqs.~(\ref{eq:cost_fm}) and (\ref{eq:cost_fc}),
          respectively,
          for the efficiency correction of the fourth and sixth order cumulants
          as a function of efficiency bins.
        }
	\label{fig:cost}
\end{figure}

Assuming that the numerical cost to calculate one mixed factorial moment
is insensitive to the order, the cost in the conventional method 
is proportional to Eq.~(\ref{eq:cost_fm}),
which grows as $\sim M^m$ for large $M$.
On the other hand, 
in the new method with Eqs.~(\ref{eq:mk_1})--(\ref{eq:mk_6}), 
the total number of cumulants, 
\begin{eqnarray}
  &&
  N_{1}^{\rm c} = 1, \quad N_{2}^{\rm c} = 3, \quad N_{3}^{\rm c} = 6,
  \nonumber \\ &&
  N_{4}^{\rm c} = 13, \quad N_{5}^{\rm c} = 24, \quad N_{6}^{\rm c} = 48, 
  \label{eq:cost_fc} 
\end{eqnarray}
is independent of $M$.
By comparing Eqs.~(\ref{eq:cost_fm}) and (\ref{eq:cost_fc}),
it is clear that the new method becomes more advantageous 
for larger $M$ especially for higher order cumulants.
In the actual numerical analyses, the cost to calculate one 
factorial moment or cumulant can grow with increasing $M$
depending on the implementation and data structure.
As this $M$ dependence can be common for both methods, 
we neglect this effect here.
In Fig.~\ref{fig:cost} we show the number of terms 
in both methods, Eqs.~(\ref{eq:cost_fm}) and (\ref{eq:cost_fc}), 
as functions of $M$ for fourth and sixth orders.

In Table~\ref{tab:cost}, we show the comparison of the CPU time to 
calculate the sixth order cumulant in both methods 
in specific implementations.
The codes are executed on the same CPU (3GHz Intel Core i7)
for $1\times10^{5}$ events and $M=4$, $8$, $12$, and $200$.
One finds that the CPU time with the conventional method 
rapidly increases as $M$ becomes larger, 
while CPU time in the new method is insensitive to $M$;
even with $M=200$, the CPU time is only about twice larger than 
$M=4$ in the new method.
This result is consistent with the cost estimate in Fig.~\ref{fig:cost}. 
Moreover, the new method is about two order 
faster than the conventional one already at $M=8$.
We, however, note that the calculation cost, of coarse, is strongly 
dependent on the implementation.

It is also notable that the new method is advantageous in simplifying 
the code and reducing momory resource.

\begin{table}[htb]
	\begin{tabular*}{60mm}{@{\extracolsep{\fill}}ccc}\hline\hline
		 $M$ & factorial moment  & new method   \\ \hline 
		 4  & $64.7$s               & $30.8$s  \\ \hline
		 8  & $17.3\times 10^{2}$s  & $31.3$s  \\ \hline 
		 12 & $14.1\times 10^{3}$s  & $32.3$s  \\ \hline
		 200 & --  & $62.7$s  \\ \hline \hline  
	\vspace{1mm}
	\end{tabular*}
	\label{tab:cost}
	\caption{Comparison of CPU time to calculate the sixth order cumulant 
	between the conventional and new methods.} 
\end{table}


\section{Two-distribution model \label{sec:ana}}

In the rest of this paper we focus on the effect of 
using the averaged efficiency for different efficiency bins.
In this section, we first consider a simple problem which can be
treated analytically.

We consider a measurement of two kinds of particle number distributions $P(N_{\rm A})$ and $P(N_{\rm B})$ 
by detectors having different efficiencies $\varepsilon_{\rm A}$ and $\varepsilon_{\rm B}$, respectively.
We assume that the two distributions are equivalent and independent, and their cumulants are given by
\begin{equation}
	\ave{N_{\rm A}^{m}}_{\rm c}=\ave{N_{\rm B}^{m}}_{\rm c}=C_{m}.
\end{equation}
We are interested in the cumulants of the total particle number $N=N_{\rm A}+N_{\rm B}$.
Due to the additive property of cumulants for independent stochastic variables~\cite{Asakawa:2015ybt}, 
cumulants of $N$ are given by 
\begin{equation}
	K_{m} \equiv \ave{N^{m}}_{\rm c} = \ave{N_{\rm A}^{m}}_{\rm c} + \ave{N_{\rm B}^{m}}_{\rm c}  = 2C_{m}. 
        \label{eq:analytical_input} 
\end{equation}
Because of the efficiency loss, the observed particle numbers $n_{\rm A}$ and $n_{\rm B}$ have different distributions 
from those of $N_{\rm A}$ and $N_{\rm B}$.
The cumulants of $n_{\rm A}$ and $n_{\rm B}$ are represented by $C_m$ 
by the inverse procedure of Eqs.~(\ref{eq:final_C1}) and 
(\ref{eq:final_C2})~\cite{Asakawa:2015ybt}. 
For the first and second orders we have
\begin{eqnarray}
  \ave{n_{X}} &=& \varepsilon_{X}C_{1},
  \label{eq:n_A1}\\
  \ave{n_{X}^{2}}_{\rm c} &=& \varepsilon_{X}^{2}C_{2} + \varepsilon_{X}(1-\varepsilon_{X})C_{1},
  \label{eq:n_A2}
\end{eqnarray}
with $X=$A and B.
By substituting Eqs.~(\ref{eq:n_A1}) and (\ref{eq:n_A2}) into 
Eqs.~(\ref{eq:mk_1}) and (\ref{eq:mk_2}) with $M=2$,
the correct value of $K_m$ is recovered.

Now, we consider a case that the efficiency correction is performed 
by regarding $n=n_{\rm A}+n_{\rm B}$ as a particle number described by a 
single distribution function measured by an averaged efficiency $\overline{\varepsilon}=(\varepsilon_{\rm A}+\varepsilon_{\rm B})/2$.
Then, the efficiency correction would be performed by substituting 
$n=n_{\rm A}+n_{\rm B}$ and $p=\overline{\varepsilon}$ into 
the result in Sec.~\ref{sec:single} such as 
Eqs.~(\ref{eq:final_C1}) and (\ref{eq:final_C2}).
For the first order, the result of this efficiency correction is 
\begin{eqnarray}
	K_1^{\rm (ave)}   &=& \ave{N_{1}} + \ave{N_{2}} = \frac{\ave{n_{1}}}{\overline{\varepsilon}} + \frac{\ave{n_{2}}}{\overline{\varepsilon}} \nonumber \\
		&=& \frac{\varepsilon_{1}C_{1}}{\overline{\varepsilon}} + \frac{\varepsilon_{2}C_{1}}{\overline{\varepsilon}} \nonumber \\
		&=& 2C_{1}.
\end{eqnarray}
Therefore, the correct cumulant Eq.~(\ref{eq:analytical_input}) is recovered to this order.
This, however, is not the case for higher order cumulants.
By denoting the deviation of the reconstructed cumulant with average efficiency
$K_m^{\rm (ave)}$ from the original one as
\begin{eqnarray}
\Delta K_{m} = K_m - K_m^{\rm (ave)} = 2C_{m} - K_m^{\rm (ave)},
\end{eqnarray}
$\Delta K_m$ is calculated to be
\begin{eqnarray}
	\Delta K_{2} &=& \frac{1}{2}\biggl(\frac{\Delta \varepsilon}{\overline{\varepsilon}}\biggr)^{2}(C_{2}-C_{1}), 
        \label{eq:DK2} \\
	\Delta K_{3} &=& \frac{3}{2}\biggl(\frac{\Delta \varepsilon}{\overline{\varepsilon}}\biggr)^{2}(C_{3}-2C_{2}+C_{1}),
        \label{eq:DK3} \\
	\Delta K_{4} &=& \frac{1}{2}\biggl(\frac{\Delta \varepsilon}{\overline{\varepsilon}}\biggr)^{2}(6C_{4}-18C_{3}+19C_{2}-7C_{1}) 
	\nonumber \\ 
	&&+ \frac{1}{8}\biggl(\frac{\Delta \varepsilon}{\overline{\varepsilon}}\biggr)^{4}(C_{4}-6C_{3}+11C_{2}-6C_{1}),
        \label{eq:DK4} 
\end{eqnarray}
with $\Delta \varepsilon=\varepsilon_{\rm A}-\varepsilon_{\rm B}$. 
The nonzero $\Delta K_m$ shows that the reconstructed cumulant 
does not agree with the original one.
These results clearly show that the use of the averaged efficiency 
gives rise to a deviation in the result of the efficiency correction.
Only for Poisson distribution ($C_{1}=C_{2}=...=C_{m}$), the deviation
vanishes.

\begin{figure}[H]
	\begin{center}
	\includegraphics[width=70mm]{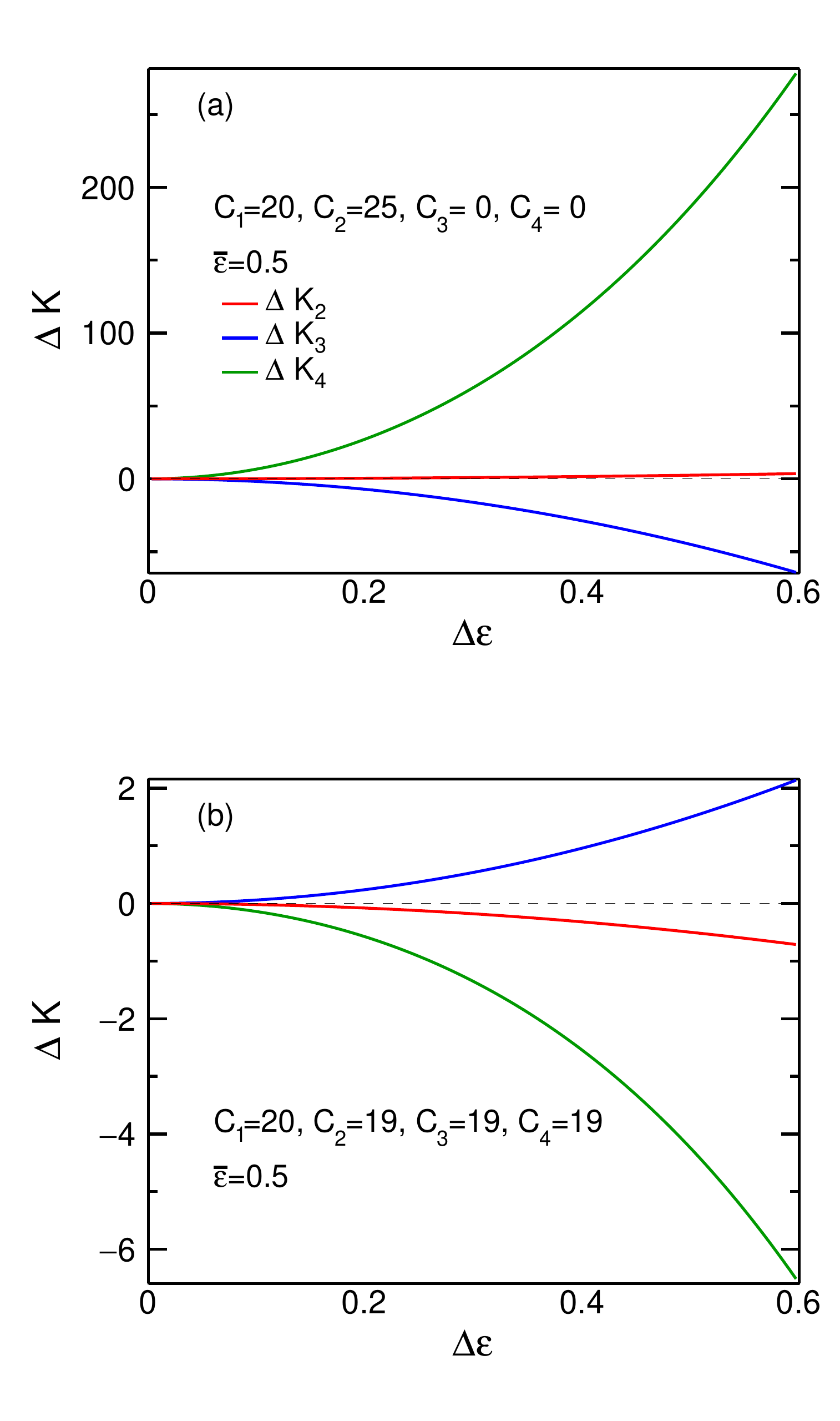}
	\end{center}
	\caption{Deviation of the efficiency corrected values of cumulants using averaged efficiency $\Delta K_m$ 
		assuming (a) Gauss distribution and (b) distribution that has 5\% smaller cumulants than Poisson distribution. }
	\label{fig:analytical}
\end{figure}

Let us see $\Delta K_m$ in specific distributions.
We first consider a Gauss distributions 
with $C_1=C_2=20$ and $C_m=0$ for $m\ge3$.
In the top panel of Fig.~\ref{fig:analytical}, 
$\Delta K_m$ with $m=2,\ 3,\ 4$ are plotted 
as functions of $\Delta \varepsilon$.
One finds that $\Delta K_m$ becomes large 
with increasing $\Delta \varepsilon$ and $m$.
Next, we consider a distribution with 
$C_1=20$ and $C_m=19$ for $m\ge2$; this distribution is 
close to Poissonian but cumulants higher than the first order 
are 5\% smaller than Poissonian values.
The $\Delta \varepsilon$ dependencies of $\Delta K_m$ in this case
is shown in the bottom panel of Fig.~\ref{fig:analytical}.
From the figure, one again obtains the same conclusion that 
$\Delta K_m$ becomes large for higher orders and larger $\Delta \varepsilon$.
These results show the importance of the use of the separated efficiencies
in the experimental analysis especially for higher orders.

\section{Numerical analysis in toy models\label{sec:toy}}

In this section, we study the effects of using averaged efficiency numerically 
in toy models by generating random events.

\subsection{Two-distribution model \label{sec:simple_toy}}

First, we analyze the two-distribution problem discussed in the previous section
numerically.
Two particle numbers $N_{\rm A}$ and $N_{\rm B}$ are independently generated 
according to Gauss distribution, 
and they are randomly sampled with the efficiencies $\varepsilon_{\rm A}$ and ${\varepsilon_{\rm B}}$ 
to obtain the measured particle numbers $n_{\rm A}$ and $n_{\rm B}$. 
We generated 100M events, and this analysis was repeated 30 times 
independently for the estimate of the statistical error.
We perform the efficiency correction by the following two methods:
\begin{enumerate}
\item
Efficiency correction with separated efficiencies for A and B.
\item
Efficiency correction using the averaged efficiency $\overline{\varepsilon}=(\varepsilon_{\rm A}+\varepsilon_{\rm B})/2$. 
\end{enumerate}
We set $C_1=20$, $C_2=25$, $\varepsilon_{\rm A}=0.3$, and 
$\varepsilon_{\rm B}=0.7$.

\begin{figure}[H]
	\begin{center}
	\includegraphics[width=70mm]{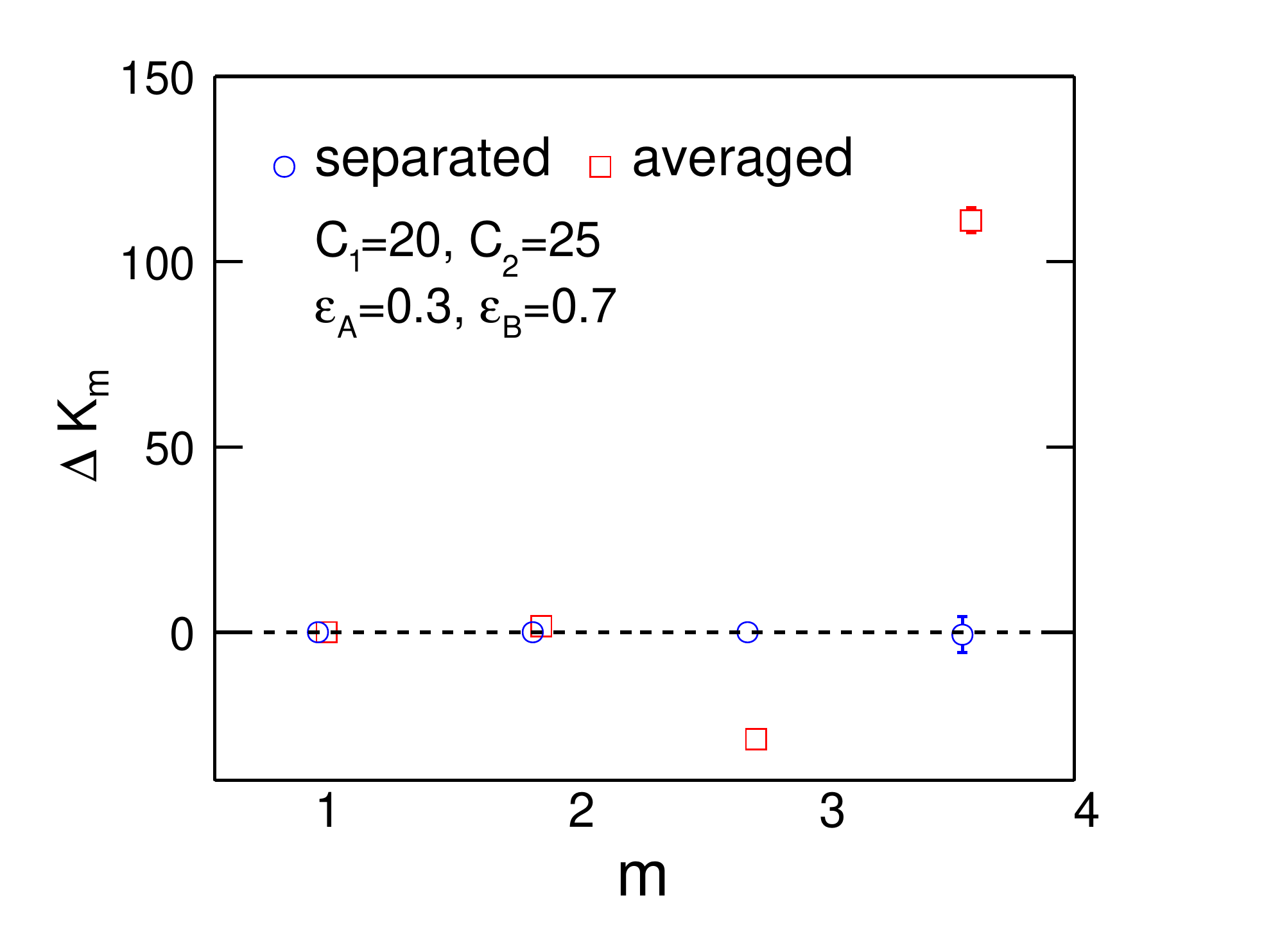}
	\end{center}
		\caption{Deviation of the efficiency corrected $m$th order cumulant from input value $\Delta K_{m}$ up to fourth order. 
		Blue circles represent the results from separated efficiencies, and red squares are results with the averaged efficiency. 
		Parameters for the distribution are $C_1=20$ and $C_2=25$. 
		Efficiencies are $\varepsilon_{A}=$0.3 and $\varepsilon_{B}=$0.7. The result can be directly compared with analytical calculation in Sec.~\ref{sec:ana}.}
	\label{fig:toy}
\end{figure}

The results of $\Delta K_{m}$ in these analyses are shown in Fig.~\ref{fig:toy} as a function of the order of the cumulant $m$. 
Blue circles are the results with separated efficiency correction. The figure shows that they reproduce correct input cumulants with $\Delta K_{m}=0$ within statistical error. 
Red squares represent results from the averaged efficiency.
The figure shows that these results give wrong
values with $\Delta K_{m}\ne0$ for $m\ge2$.
These deviations are compared with the analytic results in Eqs.~(\ref{eq:DK2})--(\ref{eq:DK4}) in Tab.~\ref{tab:comparison}.
The table shows that they are consistent with each other. 

\begin{table}[htb]
	\begin{tabular*}{80mm}{@{\extracolsep{\fill}}cccc}\hline\hline
		  $m$ &    separated   &    averaged    &   analytical     \\ \hline 
		   1  & $0\pm0.7\times10^{-3}$   &  $0\pm0.6\times10^{-3}$  & $0$      \\ 
		   2  & $0\pm0.01$   &  $1.63\pm0.01$   & $1.60$   \\
		   3  & $0\pm0.3$    &  $-28.9\pm0.2$   & $-28.8$  \\
		   4  & $-0.7\pm4.8$ &  $111.1\pm3.4$   & $115.1$  \\ \hline\hline
	\vspace{1mm}
	\end{tabular*}
	\label{tab:comparison}
	\caption{Comparison of $\Delta K_{m}$ between the numerical and the analytical calculations in the two-distribution model.}
\end{table}

\subsection{Averaged efficiencies for different particle species\label{sec:netcharge}}
In \ref{sec:ana} and \ref{sec:simple_toy}, we discussed the case with a single particle species with a unit charge. 
Next we extend the discussion to the case of the net-charge fluctuation.
In this case, we measure the charged particles without particle identifications, 
and there seems to be no problem to use averaged efficiency of charged particles for the correction.
However, when we consider the fact that the charged particles mainly consist of $\pi^{\pm}$, $K^{\pm}$ and $p^{\pm}$, 
this assumption would be violated, because those particles have different efficiencies experimentally and 
their net-particle distributions could have different probability distributions. 
Therefore, we perform a toy model analysis in order to study the effect of using the averaged efficiency assuming the net-charge distribution.
At high beam energies, one can expect that produced pions distribution is closer to the Gaussian than kaons and protons due to the large production of pions.
In this toy model, therefore, we simply set the distribution for $\pi^{\pm}$ as Gauss distribution as an extreme case, while for $K^{\pm}$ and $p^{\pm}$ as Poisson distributions.
These particles are observed with different efficiencies for different 
particle species.
These different efficiencies are used in the analysis of separated efficiency correction.
We also perform the efficiency correction with the averaged efficiencies 
for positively and negatively charged particles
\begin{equation}
	\varepsilon_{\pi Kp}^{\pm}=\frac{\sum_{i}\varepsilon_{i}^{\pm}N_{i}^{\pm}}{\sum_{i}N_{i}^{\pm}},
\end{equation}
where $i$ denotes particle species ($\pi$, $K$, $p$) and $N$ is number of produced particles.
Note that the use of the averaged efficiency for positively and negatively charged particles derives other artificial effects discussed in Ref.~\cite{tsukuba_eff_separate}.
Parameters are shown in Tab.~\ref{tab:param_netcharge}. 

\begin{table}[htb]
	\begin{tabular*}{70mm}{@{\extracolsep{\fill}}cccccc}\hline
		particles& $P(N)$ &  charge   & mean & sigma &  efficiency    \\ \hline 
		$\pi^+$  & Gauss    &  $+1$  &  30   & 8 & 0.3  \\
		$K^+$    & Poisson  &  $+1$  &  10   & --   & 0.6  \\ 
		$p$      & Poisson  &  $+1$  &  8    & --   & 0.9  \\ \hline
		$\pi^-$  & Gauss    &  $-1$  &  25   & 7 & 0.25  \\ 
		$K^-$    & Poisson  &  $-1$  &  4    & --   & 0.55 \\ 
		$\bar{p}$& Poisson  &  $-1$  &  3    & --   & 0.85  \\ \hline
	\vspace{1mm}
	\end{tabular*}
	\caption{Parameters used in the toy model discussed in \ref{sec:netcharge}.}
	\label{tab:param_netcharge}
\end{table}

\begin{figure}[H]
	\begin{center}
	\includegraphics[width=70mm]{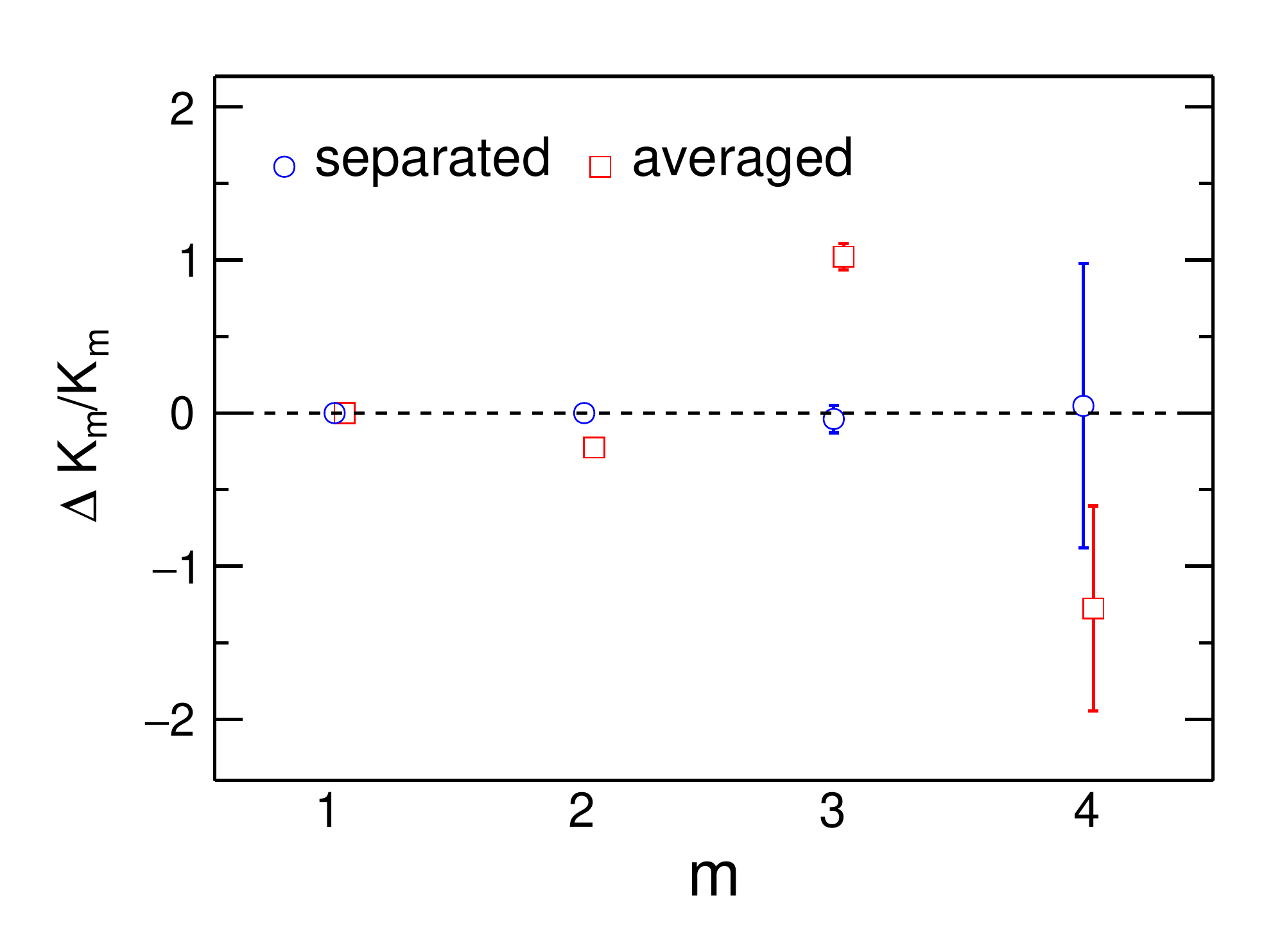}
	\end{center}
	\caption{Relative deviation of efficiency corrected $m$th order cumulant from input value ($\Delta K_{m}/K_{m}$). 
	Blue circles and red squares represent the results from separated and averaged efficiencies, respectively.}
	\label{fig:netcharge_toy}
\end{figure}

Relative deviation of efficiency corrected $m$th order cumulant from input value $\Delta K_{m}/K_{m}$ are shown in Fig.~\ref{fig:netcharge_toy} up to the fourth order. 
The figure shows that the result with the averaged efficiencies again cannot reproduce the correct value. 
Thus, we must not use averaged efficiency if there are different physics in different efficiency bins.

\subsection{Two detectors with a common source\label{sec:nonuniform}}

In current analysis for net-proton distribution at STAR, efficiency bin is divided into two $p_{T}$ regions, $0.4<p_{T}<0.8$ and $0.8<p_{T}<2.0~$GeV/c~\cite{eff_proceedings}, 
because the measurement of particles are performed in different ways for these $p_T$ regions:
Energy loss measured by Time Projection Chamber (TPC) is used for proton identification at $0.4<p_{T}<0.8~$GeV/c, 
while the mass squared measured by Time Of Flight (TOF) detector is also used at $0.8<p_{T}<2.0~$GeV/c. By including TOF detector, 
the efficiency drops at $0.8<p_{T}<2.0~$GeV/c. This $p_{T}$ dependent efficiency is implemented by dividing $p_{T}$ region at $0.8~$GeV/c.
Similarly, efficiencies would depend on $\phi$ direction.
TPC and TOF cover full azimuthal angle and have excellent particle identification capability. 
However, some of the TPC sectors are sometimes in a bad condition, which leads to the nonuniform acceptance in the $\phi$ direction. 
Let us discuss the effect by using the averaged efficiency in these conditions assuming two detectors, which may not be the case discussed in \ref{sec:netcharge}, 
because the distribution at each detector would not be determined separately. 
In other words, even if there are different kinds of particle distributions, we cannot identify those distributions at the detector level.

Setup for the toy model is as follows. Particles are randomly generated according to Gauss distributions $P(N)$, 
and let those particles randomly incident on the detector A or B with 50\% probability.
Then particles are randomly sampled by efficiencies $\varepsilon_{A}^{\pm}$ and $\varepsilon_{B}^{\pm}$. 
We apply efficiency correction on $P(N_{\rm A})$ and $P(N_{\rm B})$ with separated efficiencies or with averaged efficiency between two detectors. 
We consider the net particle number by generating charge $\pm1$ particles assuming the measurement of net-proton number cumulants.
Parameters are shown in Tab.~\ref{tab:param_nonuniform}.

\begin{table}[htb]
	\begin{tabular*}{70mm}{@{\extracolsep{\fill}}cccc||c}\hline
		$P(N)$ &  charge   & mean & sigma &  efficiency    \\ \hline 
		Gauss  &  $+1$  &  20   & $\sqrt{32}$  & $\varepsilon_{A}^{+}=0.9,\;\varepsilon_{B}^{+}=0.3$  \\ \hline
		Gauss  &  $-1$  &  8    & $\sqrt{8}$   & $\varepsilon_{A}^{-}=0.4,\;\varepsilon_{B}^{-}=0.8$  \\ \hline
		\vspace{1mm}
	\end{tabular*}
	\caption{Parameters used in the toy model discussed in \ref{sec:nonuniform}.}
	\label{tab:param_nonuniform}
\end{table}

\begin{figure}[H]
	\begin{center}
	\includegraphics[width=70mm]{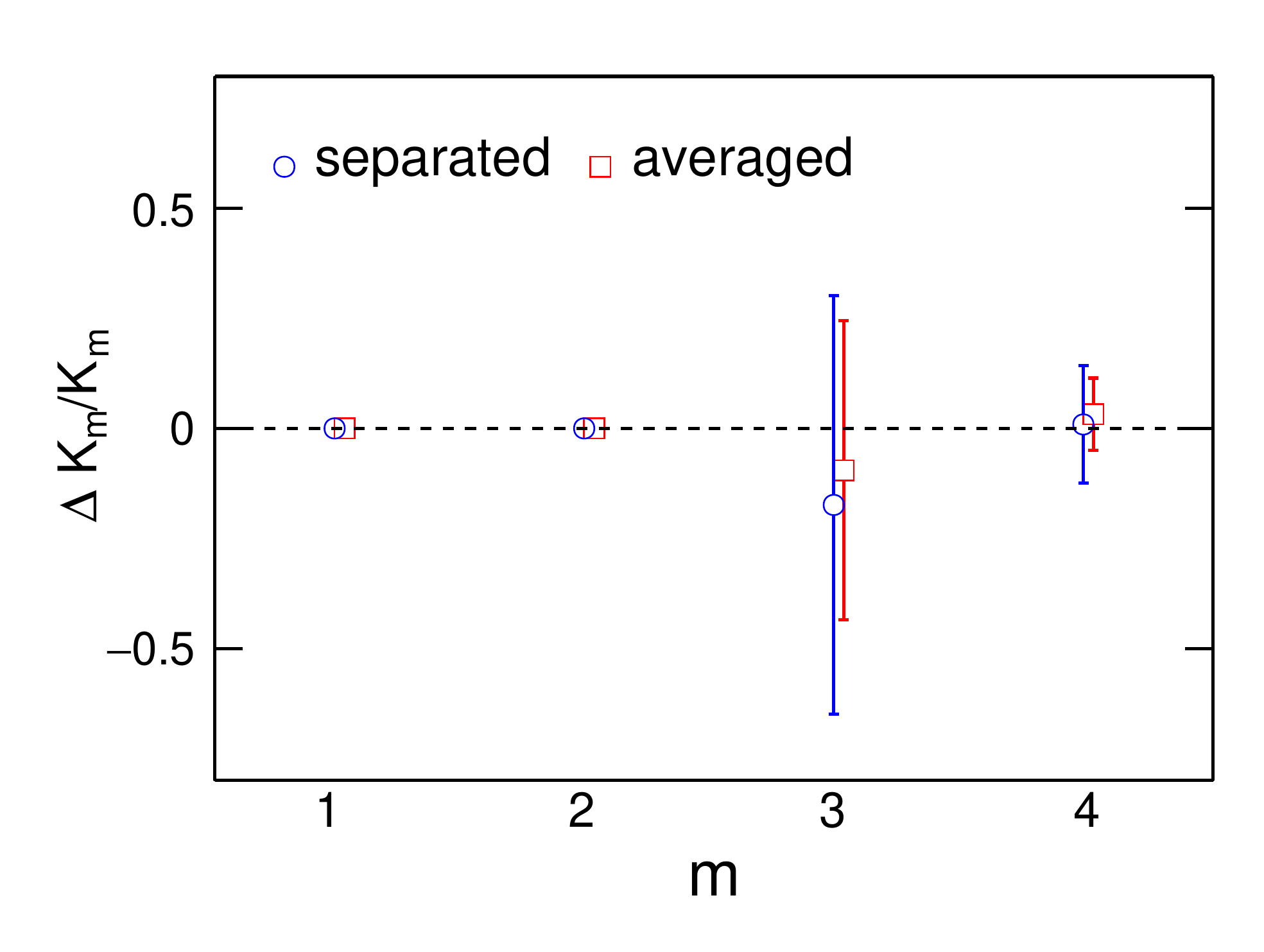}
	\end{center}
	\caption{Relative deviation of efficiency corrected $m$th order cumulant from input value $\Delta K_{m}/K_{m}$.} 
	\label{fig:nonuniform_toy}
\end{figure}

The last row in Tab.~\ref{tab:param_nonuniform} represents efficiencies that are characterized for each detector and electric charge.
Results of $\Delta K_m$ are shown in Fig.~\ref{fig:nonuniform_toy}. 
From the figure, one finds that there is no deviation for all the order of cumulants.
Note that the value of the denominator $K_{m}$ is not common
for different $m$.
This leads to the larger error for third order than fourth order
in Fig.~\ref{fig:nonuniform_toy}.
At first glance this looks strange, but we can provide a simple explanation as follows. 
When one focuses on a particle in this model, it is measured with a probability $(\varepsilon_{A}^{\pm}+\varepsilon_{B}^{\pm})/2$ randomly and independently. 
Therefore, this is exactly the case of single efficiency bin with the averaged efficiency.
This result indicates that the efficiency correction with averaged efficiency works well when underlying physics is identical for different efficiency bins.
However, for nonuniform acceptance in real experiment, 
one needs to check whether the results obtained from averaged 
efficiencies are consistent with the separated efficiencies.

\section{Summary}

In this paper, we derived formulas for the efficiency correction
with many efficiency bins.
In our method, the formulas are obtained easily compared to 
Ref.~\cite{psd_Kitazawa}, but the numerical cost is 
drastically reduced compared to Refs.~\cite{eff_psd_volker,eff_xiaofeng} 
when the number of efficiency bins and order of the cumulant are large.
The efficiency correction for higher order cumulants with many bins
thus can be carried out effectively in our method.
The result is then applied to the efficiency correction in simple models
to study the effect of using averaged efficiency in Secs.~\ref{sec:ana} and \ref{sec:toy}.
We have shown that the use of the averaged efficiency can lead to 
wrong corrected values if underlying physics is different in efficiency bins.
This result indicates that separated efficiencies have to be used to perform the efficiency correction correctly.
For example, it would be important to take account of the nonuniform acceptance along azimuthal angle 
and the $p_T$ dependencies of efficiency for the accurate efficiency correction.

Final remarks are in order.
First, although we used the binomial model throughout this paper,
this model is justified only when the efficiencies for individual 
particles are independent \cite{Asakawa:2015ybt}.
When the correlations between individual particles are not negligible,
these effects have to be considered \cite{binomial_breaking}.
Second, experimental analyses usually measure proton number 
cumulants as proxies of baryon number cumulants.
In Refs.~\cite{Kitazawa:2011wh,eff_kitazawa}, it is shown that 
the measurement of protons corresponds to the measurement of baryons
with $50\%$ efficiency loss. Therefore, the baryon number cumulants
can in principle be constructed from those of protons using 
efficiency correction. In this case, the use of the binomial model
is justified owing to isospin randomization \cite{Kitazawa:2011wh}.

\section{Acknowledgement}

The authors thank X.~Luo and N.~Xu for useful discussions.
T.~N. thanks P.~Tribedy for the idea of the toy model discussed in Sec.~\ref{sec:netcharge}. 
M.~K. thanks stimulating discussions in the INT program ``Exploring the QCD Phase Diagram through Energy Scans'', Seattle, Sep.~19 -- Oct.~14, especially A.~Bzdak and V.~Koch.
We acknowledge support from MEXT and JSPS KAKENHI Grant Number 25105504 and Super Global University Program in University of Tsukuba.


\appendix
\section{Net-particle in simple case}
\label{sec:net}

In the case of net-particle with single efficiency bin, explicit formulas for efficiency correction 
can be derived from Eqs.~(\ref{eq:mk_1})--(\ref{eq:multi_q}) 
by substituting $M=2$, $a_{1}=1$, $a_{2}=-1$, and $p_{1}=p_{2}=p$.
By defining $n_{\rm net}=n_{1}-n_{2}$ and $n_{\rm tot}=n_{1}+n_{2}$, 
the formulas up to sixth order are given by
\begin{widetext}
\begin{eqnarray}
	\ave{Q}_{\rm c} &=& \frac{1}{p}\ave{n_{\rm net}}_{\rm c}, \\
	\ave{Q^{2}}_{\rm c} &=& \frac{1}{p^{2}}\ave{n_{\rm net}^{2}}_{\rm c} 
				+ \Bigl(-\frac{1}{p^{2}}+\frac{1}{p}\Bigr)\ave{n_{\rm tot}}, \\
	\ave{Q^{3}}_{\rm c} &=& \frac{1}{p^{3}}\ave{n_{\rm net}^{3}}_{\rm c} 
				+ \Bigl(-\frac{3}{p^{3}}+\frac{3}{p^{2}}\Bigr)\ave{n_{\rm net}n_{\rm tot}}_{\rm c}
				+ \Bigl(\frac{2}{p^{3}}-\frac{3}{p^{2}}+\frac{1}{p}\Bigr)\ave{n_{\rm net}}_{\rm c}, \\
	\ave{Q^{4}}_{\rm c} &=& \frac{1}{p^{4}}\ave{n_{\rm net}^{4}}_{\rm c}
				+ \Bigl(-\frac{6}{p^{4}}+\frac{6}{p^{3}}\Bigr)\ave{n_{\rm net}^{2}n_{\rm tot}}_{\rm c}
				+ \Bigl(\frac{8}{p^{4}}-\frac{12}{p^{3}}+\frac{4}{p^{2}}\Bigr)\ave{n_{\rm net}^{2}}_{\rm c}
				+ \Bigl(\frac{3}{p^{4}}-\frac{6}{p^{3}}+\frac{3}{p^{2}}\Bigr)\ave{n_{\rm tot}^{2}}_{\rm c} \nonumber \\
		             && + \Bigl(-\frac{6}{p^{4}}+\frac{12}{p^{3}}-\frac{7}{p^{2}}+\frac{1}{p}\Bigr)\ave{n_{\rm tot}}_{\rm c}, \\
	\ave{Q^{5}}_{\rm c} &=& \frac{1}{p^{5}}\ave{n_{\rm net}^{5}}_{\rm c}
				+ \Bigl(-\frac{10}{p^{5}}+\frac{10}{p^{4}}\Bigr)\ave{n_{\rm net}^{3}n_{\rm tot}}_{\rm c}
				+ \Bigl(\frac{20}{p^{5}}-\frac{30}{p^{4}}+\frac{10}{p^{3}}\Bigr)\ave{n_{\rm net}^{3}}_{\rm c}
				+ \Bigl(\frac{15}{p^{5}}-\frac{30}{p^{4}}+\frac{15}{p^{3}}\Bigr)\ave{n_{\rm net}n_{\rm tot}^{2}}_{\rm c} \nonumber \\
			     && + \Bigl(-\frac{50}{p^{5}}+\frac{110}{p^{4}}-\frac{75}{p^{3}}+\frac{15}{p^{2}}\Bigr)\ave{n_{\rm net}n_{\rm tot}}_{\rm c}
				+ \Bigl(\frac{24}{p^{5}}-\frac{60}{p^{4}}+\frac{50}{p^{3}}-\frac{15}{p^{2}}+\frac{1}{p}\Bigr)\ave{n_{\rm net}}_{\rm c}, \\
	\ave{Q^{6}}_{\rm c} &=& \frac{1}{p^{6}}\ave{n_{\rm net}^{6}}_{\rm c}
				+ \Bigl(-\frac{15}{p^{6}}+\frac{15}{p^{5}}\Bigr)\ave{n_{\rm net}^{4}n_{\rm tot}}_{\rm c}
				+ \Bigl(\frac{40}{p^{6}}-\frac{60}{p^{5}}+\frac{20}{p^{4}}\Bigr)\ave{n_{\rm net}^{4}}_{\rm c} \nonumber \\
			     && + \Bigl(\frac{45}{p^{6}}-\frac{90}{p^{5}}+\frac{45}{p^{4}}\Bigr)\ave{n_{\rm net}^{2}n_{\rm tot}^{2}}_{\rm c}
			        + \Bigl(-\frac{15}{p^{6}}+\frac{45}{p^{5}}-\frac{45}{p^{4}}+\frac{15}{p^{3}}\Bigr)\ave{n_{\rm tot}^{3}}_{\rm c}
				+ \Bigl(-\frac{210}{p^{6}}+\frac{480}{p^{5}}-\frac{345}{p^{4}}+\frac{75}{p^{3}}\Bigr)\ave{n_{\rm net}^{2}n_{\rm tot}}_{\rm c} \nonumber \\
				&& + \Bigl(\frac{184}{p^{6}}-\frac{480}{p^{5}}+\frac{430}{p^{4}}-\frac{150}{p^{3}}+\frac{16}{p^{2}}\Bigr)\ave{n_{\rm net}^{2}}_{\rm c}
			        + \Bigl(\frac{90}{p^{6}}-\frac{270}{p^{5}}+\frac{285}{p^{4}}-\frac{120}{p^{3}}+\frac{15}{p^{2}}\Bigr)\ave{n_{\rm tot}^{2}}_{\rm c} \nonumber \\
			     && + \Bigl(-\frac{120}{p^{6}}+\frac{360}{p^{5}}-\frac{390}{p^{4}}+\frac{180}{p^{3}}-\frac{31}{p^{2}}+\frac{1}{p}\Bigr)\ave{n_{\rm tot}}_{\rm c} .
\end{eqnarray}
\end{widetext}


\bibliography{main}

\providecommand{\noopsort}[1]{}\providecommand{\singleletter}[1]{#1}%
\begin{thebibliography}{21}%
\makeatletter
\providecommand \@ifxundefined [1]{%
 \@ifx{#1\undefined}
}%
\providecommand \@ifnum [1]{%
 \ifnum #1\expandafter \@firstoftwo
 \else \expandafter \@secondoftwo
 \fi
}%
\providecommand \@ifx [1]{%
 \ifx #1\expandafter \@firstoftwo
 \else \expandafter \@secondoftwo
 \fi
}%
\providecommand \natexlab [1]{#1}%
\providecommand \enquote  [1]{``#1''}%
\providecommand \bibnamefont  [1]{#1}%
\providecommand \bibfnamefont [1]{#1}%
\providecommand \citenamefont [1]{#1}%
\providecommand \href@noop [0]{\@secondoftwo}%
\providecommand \href [0]{\begingroup \@sanitize@url \@href}%
\providecommand \@href[1]{\@@startlink{#1}\@@href}%
\providecommand \@@href[1]{\endgroup#1\@@endlink}%
\providecommand \@sanitize@url [0]{\catcode `\\12\catcode `\$12\catcode
  `\&12\catcode `\#12\catcode `\^12\catcode `\_12\catcode `\%12\relax}%
\providecommand \@@startlink[1]{}%
\providecommand \@@endlink[0]{}%
\providecommand \url  [0]{\begingroup\@sanitize@url \@url }%
\providecommand \@url [1]{\endgroup\@href {#1}{\urlprefix }}%
\providecommand \urlprefix  [0]{URL }%
\providecommand \Eprint [0]{\href }%
\providecommand \doibase [0]{http://dx.doi.org/}%
\providecommand \selectlanguage [0]{\@gobble}%
\providecommand \bibinfo  [0]{\@secondoftwo}%
\providecommand \bibfield  [0]{\@secondoftwo}%
\providecommand \translation [1]{[#1]}%
\providecommand \BibitemOpen [0]{}%
\providecommand \bibitemStop [0]{}%
\providecommand \bibitemNoStop [0]{.\EOS\space}%
\providecommand \EOS [0]{\spacefactor3000\relax}%
\providecommand \BibitemShut  [1]{\csname bibitem#1\endcsname}%
\let\auto@bib@innerbib\@empty
\bibitem [{\citenamefont {Asakawa}\ \emph {et~al.}(2000)\citenamefont
  {Asakawa}, \citenamefont {Heinz},\ and\ \citenamefont
  {Muller}}]{fluctuation}%
  \BibitemOpen
  \bibfield  {author} {\bibinfo {author} {\bibfnamefont {M.}~\bibnamefont
  {Asakawa}}, \bibinfo {author} {\bibfnamefont {U.~W.}\ \bibnamefont {Heinz}},
  \ and\ \bibinfo {author} {\bibfnamefont {B.}~\bibnamefont {Muller}},\ }\href
  {\doibase 10.1103/PhysRevLett.85.2072} {\bibfield  {journal} {\bibinfo
  {journal} {Phys. Rev. Lett.}\ }\textbf {\bibinfo {volume} {85}},\ \bibinfo
  {pages} {2072} (\bibinfo {year} {2000})},\ \Eprint
  {http://arxiv.org/abs/hep-ph/0003169} {arXiv:hep-ph/0003169 [hep-ph]}
  \BibitemShut {NoStop}%
\bibitem [{\citenamefont {Jeon}\ and\ \citenamefont
  {Koch}(2000)}]{Jeon:2000wg}%
  \BibitemOpen
  \bibfield  {author} {\bibinfo {author} {\bibfnamefont {S.}~\bibnamefont
  {Jeon}}\ and\ \bibinfo {author} {\bibfnamefont {V.}~\bibnamefont {Koch}},\
  }\href {\doibase 10.1103/PhysRevLett.85.2076} {\bibfield  {journal} {\bibinfo
   {journal} {Phys. Rev. Lett.}\ }\textbf {\bibinfo {volume} {85}},\ \bibinfo
  {pages} {2076} (\bibinfo {year} {2000})},\ \Eprint
  {http://arxiv.org/abs/hep-ph/0003168} {arXiv:hep-ph/0003168 [hep-ph]}
  \BibitemShut {NoStop}%
\bibitem [{\citenamefont {Ejiri}\ \emph {et~al.}(2006)\citenamefont {Ejiri},
  \citenamefont {Karsch},\ and\ \citenamefont {Redlich}}]{susceptibility}%
  \BibitemOpen
  \bibfield  {author} {\bibinfo {author} {\bibfnamefont {S.}~\bibnamefont
  {Ejiri}}, \bibinfo {author} {\bibfnamefont {F.}~\bibnamefont {Karsch}}, \
  and\ \bibinfo {author} {\bibfnamefont {K.}~\bibnamefont {Redlich}},\ }\href
  {\doibase 10.1016/j.physletb.2005.11.083} {\bibfield  {journal} {\bibinfo
  {journal} {Phys. Lett.}\ }\textbf {\bibinfo {volume} {B633}},\ \bibinfo
  {pages} {275} (\bibinfo {year} {2006})},\ \Eprint
  {http://arxiv.org/abs/hep-ph/0509051} {arXiv:hep-ph/0509051 [hep-ph]}
  \BibitemShut {NoStop}%
\bibitem [{\citenamefont {Stephanov}(2009)}]{correlation}%
  \BibitemOpen
  \bibfield  {author} {\bibinfo {author} {\bibfnamefont {M.~A.}\ \bibnamefont
  {Stephanov}},\ }\href {\doibase 10.1103/PhysRevLett.102.032301} {\bibfield
  {journal} {\bibinfo  {journal} {Phys. Rev. Lett.}\ }\textbf {\bibinfo
  {volume} {102}},\ \bibinfo {pages} {032301} (\bibinfo {year} {2009})},\
  \Eprint {http://arxiv.org/abs/0809.3450} {arXiv:0809.3450 [hep-ph]}
  \BibitemShut {NoStop}%
\bibitem [{\citenamefont {Asakawa}\ \emph {et~al.}(2009)\citenamefont
  {Asakawa}, \citenamefont {Ejiri},\ and\ \citenamefont
  {Kitazawa}}]{Asakawa:2009aj}%
  \BibitemOpen
  \bibfield  {author} {\bibinfo {author} {\bibfnamefont {M.}~\bibnamefont
  {Asakawa}}, \bibinfo {author} {\bibfnamefont {S.}~\bibnamefont {Ejiri}}, \
  and\ \bibinfo {author} {\bibfnamefont {M.}~\bibnamefont {Kitazawa}},\ }\href
  {\doibase 10.1103/PhysRevLett.103.262301} {\bibfield  {journal} {\bibinfo
  {journal} {Phys. Rev. Lett.}\ }\textbf {\bibinfo {volume} {103}},\ \bibinfo
  {pages} {262301} (\bibinfo {year} {2009})},\ \Eprint
  {http://arxiv.org/abs/0904.2089} {arXiv:0904.2089 [nucl-th]} \BibitemShut
  {NoStop}%
\bibitem [{\citenamefont {Friman}\ \emph {et~al.}(2011)\citenamefont {Friman},
  \citenamefont {Karsch}, \citenamefont {Redlich},\ and\ \citenamefont
  {Skokov}}]{Friman}%
  \BibitemOpen
  \bibfield  {author} {\bibinfo {author} {\bibfnamefont {B.}~\bibnamefont
  {Friman}}, \bibinfo {author} {\bibfnamefont {F.}~\bibnamefont {Karsch}},
  \bibinfo {author} {\bibfnamefont {K.}~\bibnamefont {Redlich}}, \ and\
  \bibinfo {author} {\bibfnamefont {V.}~\bibnamefont {Skokov}},\ }\href
  {\doibase 10.1140/epjc/s10052-011-1694-2} {\bibfield  {journal} {\bibinfo
  {journal} {Eur. Phys. J.}\ }\textbf {\bibinfo {volume} {C71}},\ \bibinfo
  {pages} {1694} (\bibinfo {year} {2011})},\ \Eprint
  {http://arxiv.org/abs/1103.3511} {arXiv:1103.3511 [hep-ph]} \BibitemShut
  {NoStop}%
\bibitem [{\citenamefont {Asakawa}\ and\ \citenamefont
  {Kitazawa}(2016)}]{Asakawa:2015ybt}%
  \BibitemOpen
  \bibfield  {author} {\bibinfo {author} {\bibfnamefont {M.}~\bibnamefont
  {Asakawa}}\ and\ \bibinfo {author} {\bibfnamefont {M.}~\bibnamefont
  {Kitazawa}},\ }\href {\doibase 10.1016/j.ppnp.2016.04.002} {\bibfield
  {journal} {\bibinfo  {journal} {Prog. Part. Nucl. Phys.}\ }\textbf {\bibinfo
  {volume} {90}},\ \bibinfo {pages} {299} (\bibinfo {year} {2016})},\ \Eprint
  {http://arxiv.org/abs/1512.05038} {arXiv:1512.05038 [nucl-th]} \BibitemShut
  {NoStop}%
\bibitem [{\citenamefont {Luo}\ and\ \citenamefont {Xu}(2017)}]{Luo:2017faz}%
  \BibitemOpen
  \bibfield  {author} {\bibinfo {author} {\bibfnamefont {X.}~\bibnamefont
  {Luo}}\ and\ \bibinfo {author} {\bibfnamefont {N.}~\bibnamefont {Xu}},\
  }\href@noop {} {\  (\bibinfo {year} {2017})},\ \Eprint
  {http://arxiv.org/abs/1701.02105} {arXiv:1701.02105 [nucl-ex]} \BibitemShut
  {NoStop}%
\bibitem [{\citenamefont {Adamczyk}\ \emph
  {et~al.}(2014{\natexlab{a}})\citenamefont {Adamczyk} \emph
  {et~al.}}]{net_proton}%
  \BibitemOpen
  \bibfield  {author} {\bibinfo {author} {\bibfnamefont {L.}~\bibnamefont
  {Adamczyk}} \emph {et~al.} (\bibinfo {collaboration} {STAR}),\ }\href
  {\doibase 10.1103/PhysRevLett.112.032302} {\bibfield  {journal} {\bibinfo
  {journal} {Phys. Rev. Lett.}\ }\textbf {\bibinfo {volume} {112}},\ \bibinfo
  {pages} {032302} (\bibinfo {year} {2014}{\natexlab{a}})},\ \Eprint
  {http://arxiv.org/abs/1309.5681} {arXiv:1309.5681 [nucl-ex]} \BibitemShut
  {NoStop}%
\bibitem [{\citenamefont {Adamczyk}\ \emph
  {et~al.}(2014{\natexlab{b}})\citenamefont {Adamczyk} \emph
  {et~al.}}]{net_charge}%
  \BibitemOpen
  \bibfield  {author} {\bibinfo {author} {\bibfnamefont {L.}~\bibnamefont
  {Adamczyk}} \emph {et~al.} (\bibinfo {collaboration} {STAR}),\ }\href
  {\doibase 10.1103/PhysRevLett.113.092301} {\bibfield  {journal} {\bibinfo
  {journal} {Phys. Rev. Lett.}\ }\textbf {\bibinfo {volume} {113}},\ \bibinfo
  {pages} {092301} (\bibinfo {year} {2014}{\natexlab{b}})},\ \Eprint
  {http://arxiv.org/abs/1402.1558} {arXiv:1402.1558 [nucl-ex]} \BibitemShut
  {NoStop}%
\bibitem [{\citenamefont {Chen}(2013)}]{Lizhu}%
  \BibitemOpen
  \bibfield  {author} {\bibinfo {author} {\bibfnamefont {L.}~\bibnamefont
  {Chen}} (\bibinfo {collaboration} {STAR}),\ }\bibfield  {booktitle} {\emph
  {\bibinfo {booktitle} {{Proceedings, 23rd International Conference on
  Ultrarelativistic Nucleus-Nucleus Collisions : Quark Matter 2012 (QM 2012):
  Washington, DC, USA, August 13-18, 2012}}},\ }\href {\doibase
  10.1016/j.nuclphysa.2013.02.051} {\bibfield  {journal} {\bibinfo  {journal}
  {Nucl. Phys.}\ }\textbf {\bibinfo {volume} {A904-905}},\ \bibinfo {pages}
  {471c} (\bibinfo {year} {2013})}\BibitemShut {NoStop}%
\bibitem [{\citenamefont {Kitazawa}\ and\ \citenamefont
  {Asakawa}(2012{\natexlab{a}})}]{eff_kitazawa}%
  \BibitemOpen
  \bibfield  {author} {\bibinfo {author} {\bibfnamefont {M.}~\bibnamefont
  {Kitazawa}}\ and\ \bibinfo {author} {\bibfnamefont {M.}~\bibnamefont
  {Asakawa}},\ }\href {\doibase 10.1103/PhysRevC.86.024904,
  10.1103/PhysRevC.86.069902} {\bibfield  {journal} {\bibinfo  {journal} {Phys.
  Rev.}\ }\textbf {\bibinfo {volume} {C86}},\ \bibinfo {pages} {024904}
  (\bibinfo {year} {2012}{\natexlab{a}})},\ \bibinfo {note} {[Erratum: Phys.
  Rev.C86,069902(2012)]},\ \Eprint {http://arxiv.org/abs/1205.3292}
  {arXiv:1205.3292 [nucl-th]} \BibitemShut {NoStop}%
\bibitem [{\citenamefont {Bzdak}\ and\ \citenamefont {Koch}(2012)}]{eff_koch}%
  \BibitemOpen
  \bibfield  {author} {\bibinfo {author} {\bibfnamefont {A.}~\bibnamefont
  {Bzdak}}\ and\ \bibinfo {author} {\bibfnamefont {V.}~\bibnamefont {Koch}},\
  }\href {\doibase 10.1103/PhysRevC.86.044904} {\bibfield  {journal} {\bibinfo
  {journal} {Phys. Rev.}\ }\textbf {\bibinfo {volume} {C86}},\ \bibinfo {pages}
  {044904} (\bibinfo {year} {2012})},\ \Eprint {http://arxiv.org/abs/1206.4286}
  {arXiv:1206.4286 [nucl-th]} \BibitemShut {NoStop}%
\bibitem [{\citenamefont {Bzdak}\ and\ \citenamefont
  {Koch}(2015)}]{eff_psd_volker}%
  \BibitemOpen
  \bibfield  {author} {\bibinfo {author} {\bibfnamefont {A.}~\bibnamefont
  {Bzdak}}\ and\ \bibinfo {author} {\bibfnamefont {V.}~\bibnamefont {Koch}},\
  }\href {\doibase 10.1103/PhysRevC.91.027901} {\bibfield  {journal} {\bibinfo
  {journal} {Phys. Rev.}\ }\textbf {\bibinfo {volume} {C91}},\ \bibinfo {pages}
  {027901} (\bibinfo {year} {2015})},\ \Eprint {http://arxiv.org/abs/1312.4574}
  {arXiv:1312.4574 [nucl-th]} \BibitemShut {NoStop}%
\bibitem [{\citenamefont {Luo}(2015)}]{eff_xiaofeng}%
  \BibitemOpen
  \bibfield  {author} {\bibinfo {author} {\bibfnamefont {X.}~\bibnamefont
  {Luo}},\ }\href {\doibase 10.1103/PhysRevC.91.034907} {\bibfield  {journal}
  {\bibinfo  {journal} {Phys. Rev.}\ }\textbf {\bibinfo {volume} {C91}},\
  \bibinfo {pages} {034907} (\bibinfo {year} {2015})},\ \Eprint
  {http://arxiv.org/abs/1410.3914} {arXiv:1410.3914 [physics.data-an]}
  \BibitemShut {NoStop}%
\bibitem [{\citenamefont {Kitazawa}(2016)}]{psd_Kitazawa}%
  \BibitemOpen
  \bibfield  {author} {\bibinfo {author} {\bibfnamefont {M.}~\bibnamefont
  {Kitazawa}},\ }\href {\doibase 10.1103/PhysRevC.93.044911} {\bibfield
  {journal} {\bibinfo  {journal} {Phys. Rev.}\ }\textbf {\bibinfo {volume}
  {C93}},\ \bibinfo {pages} {044911} (\bibinfo {year} {2016})},\ \Eprint
  {http://arxiv.org/abs/1602.01234} {arXiv:1602.01234 [nucl-th]} \BibitemShut
  {NoStop}%
\bibitem [{\citenamefont {Bzdak}\ \emph {et~al.}(2016)\citenamefont {Bzdak},
  \citenamefont {Holzmann},\ and\ \citenamefont {Koch}}]{binomial_breaking}%
  \BibitemOpen
  \bibfield  {author} {\bibinfo {author} {\bibfnamefont {A.}~\bibnamefont
  {Bzdak}}, \bibinfo {author} {\bibfnamefont {R.}~\bibnamefont {Holzmann}}, \
  and\ \bibinfo {author} {\bibfnamefont {V.}~\bibnamefont {Koch}},\ }\href
  {\doibase 10.1103/PhysRevC.94.064907} {\bibfield  {journal} {\bibinfo
  {journal} {Phys. Rev.}\ }\textbf {\bibinfo {volume} {C94}},\ \bibinfo {pages}
  {064907} (\bibinfo {year} {2016})},\ \Eprint
  {http://arxiv.org/abs/1603.09057} {arXiv:1603.09057 [nucl-th]} \BibitemShut
  {NoStop}%
\bibitem [{\citenamefont {Nonaka}\ \emph {et~al.}(2016)\citenamefont {Nonaka},
  \citenamefont {Sugiura}, \citenamefont {Esumi}, \citenamefont {Masui},\ and\
  \citenamefont {Luo}}]{tsukuba_eff_separate}%
  \BibitemOpen
  \bibfield  {author} {\bibinfo {author} {\bibfnamefont {T.}~\bibnamefont
  {Nonaka}}, \bibinfo {author} {\bibfnamefont {T.}~\bibnamefont {Sugiura}},
  \bibinfo {author} {\bibfnamefont {S.}~\bibnamefont {Esumi}}, \bibinfo
  {author} {\bibfnamefont {H.}~\bibnamefont {Masui}}, \ and\ \bibinfo {author}
  {\bibfnamefont {X.}~\bibnamefont {Luo}},\ }\href {\doibase
  10.1103/PhysRevC.94.034909} {\bibfield  {journal} {\bibinfo  {journal} {Phys.
  Rev.}\ }\textbf {\bibinfo {volume} {C94}},\ \bibinfo {pages} {034909}
  (\bibinfo {year} {2016})},\ \Eprint {http://arxiv.org/abs/1604.06212}
  {arXiv:1604.06212 [nucl-th]} \BibitemShut {NoStop}%
\bibitem [{\citenamefont {Kitazawa}\ and\ \citenamefont
  {Luo}(2017)}]{Kitazawa:2017ljq}%
  \BibitemOpen
  \bibfield  {author} {\bibinfo {author} {\bibfnamefont {M.}~\bibnamefont
  {Kitazawa}}\ and\ \bibinfo {author} {\bibfnamefont {X.}~\bibnamefont {Luo}},\
  }\href@noop {} {\  (\bibinfo {year} {2017})},\ \Eprint
  {http://arxiv.org/abs/1704.04909} {arXiv:1704.04909 [nucl-th]} \BibitemShut
  {NoStop}%
\bibitem [{eff(2015)}]{eff_proceedings}%
  \BibitemOpen
  \href@noop {} {\emph {\bibinfo {title} {{Proceedings, 9th International
  Workshop on Critical Point and Onset of Deconfinement (CPOD 2014)}}}},\ Vol.\
  \bibinfo {volume} {CPOD2014}\ (\bibinfo {year} {2015})\ \Eprint
  {http://arxiv.org/abs/1503.02558} {arXiv:1503.02558 [nucl-ex]} \BibitemShut
  {NoStop}%
\bibitem [{\citenamefont {Kitazawa}\ and\ \citenamefont
  {Asakawa}(2012{\natexlab{b}})}]{Kitazawa:2011wh}%
  \BibitemOpen
  \bibfield  {author} {\bibinfo {author} {\bibfnamefont {M.}~\bibnamefont
  {Kitazawa}}\ and\ \bibinfo {author} {\bibfnamefont {M.}~\bibnamefont
  {Asakawa}},\ }\href {\doibase 10.1103/PhysRevC.85.021901} {\bibfield
  {journal} {\bibinfo  {journal} {Phys. Rev.}\ }\textbf {\bibinfo {volume}
  {C85}},\ \bibinfo {pages} {021901} (\bibinfo {year} {2012}{\natexlab{b}})},\
  \Eprint {http://arxiv.org/abs/1107.2755} {arXiv:1107.2755 [nucl-th]}
  \BibitemShut {NoStop}%
\end{thebibliography}%

\end{document}